\documentclass{mn2e}
\input psfig.sty

\newcommand{\ledd}{L_{\rm E}}
\newcommand{\nh}{N_{\rm H}}

\newcommand{\ee}{$e^\pm$}
\newcommand{\g}{$\gamma$}
\newcommand{\lh}{\ell_{\rm h}}
\newcommand{\ls}{\ell_{\rm s}}
\newcommand{\lth}{\ell_{\rm th}}
\newcommand{\lnth}{\ell_{\rm nth}}

\newcommand{\source}{GRS 1915+105}
\newcommand{\msun}{{\rm M}_{\sun}}
\newcommand{\xte}{{\it RXTE}}

\newcommand{\gro}{{\it CGRO}}

\newcommand{\gmin}{\gamma_{\rm min}}
\newcommand{\gmax}{\gamma_{\rm max}}
\newcommand{\cnu}{\chi^2_\nu}

\newcommand{\apj}{ApJ}

\newcommand{\aap}{A\&A}

\title{GRS 1915+105: the distance, radiative processes and energy-dependent variability}

\author[A. A. Zdziarski et al.]
{Andrzej A. Zdziarski,$^1$\thanks{E-mail:
aaz@camk.edu.pl} Marek Gierli\'nski,$^{2,3}$\thanks{E-mail: Marek.Gierlinski@durham.ac.uk} A. R. Rao,$^4$ S. V. Vadawale,$^{5,4}$\newauthor and Joanna Miko{\l}ajewska$^1$\\
$^1$Centrum Astronomiczne im.\ M. Kopernika, Bartycka 18, 00-716 Warszawa, Poland\\
$^2$Department of Physics, University of Durham, Durham DH1~3LE, UK\\
$^3$Obserwatorium Astronomiczne Uniwersytetu Jagiello\'nskiego, Orla 171, 30-244 Krak{\'o}w, Poland\\
$^4$Tata Institute of Fundamental Research, Homi Bhabha Road,
Bombay 400 005, India\\
$^5$Harvard-Smithsonian Center for Astrophysics, 60 Garden St., MA 02138, USA\\
}

\date{Accepted 2005 March 31. Received 2005 January 13}

\pagerange{1--14}
\pubyear{2005}

\begin{document}

\maketitle

\label{firstpage}

\begin{abstract}  
We present an exhaustive analysis of five broad-band observations of GRS 1915+105 in two variability states, $\chi$ and $\omega$, observed simultaneously by the PCA and HEXTE detectors aboard {\it RXTE\/} and the OSSE detector aboard {\it CGRO}. We find all the spectra well fitted by Comptonization of disc blackbody photons, with very strong evidence for the presence of a nonthermal electron component in the Comptonizing plasma. Both the energy and the power spectra in the $\chi$ state are typical to the very high/intermediate state of black-hole binaries. The spectrum of the $\omega$ state is characterized by a strong blackbody component Comptonized by thermal electrons and a weak nonthermal tail.  We then calculate rms spectra (fractional variability as functions of energy) for the PCA data. We accurately model the rms spectra by coherent superposition of variability in the components implied by the spectral fits, namely a less variable blackbody and more variable Comptonization. The latter dominates at high energies, resulting in a flattening of the rms at high energies in most of the data. This is also the case for the spectra of the QPOs present in the $\chi$ state. Then, some of our data require a radial dependence of the rms of the disc blackbody. We also study the distance to the source, and find $d\simeq 11$ kpc as the most likely value, contrary to a recent claim of a much lower value.
\end{abstract}
\begin{keywords}
accretion, accretion discs -- binaries: general --  black hole
physics -- radiation mechanisms: non-thermal -- stars: individual (GRS
1915+105) -- X-rays: stars.
\end{keywords}

\section{INTRODUCTION}
\label{intro}

\source\ is a binary system consisting of an early-type K giant (Greiner et al.\ 2001b) accreting via a Roche-lobe overflow onto a black hole with the mass function of $9.5\pm 3\msun$ (Greiner, Cuby \& McCaughrean 2001a). The recent measurement of the rotational broadening of the giant combined with the orbital parameters has resulted in the determination of the black-hole and donor mass of $14\pm 4\msun$ and $0.8\pm 0.5\msun$, respectively (Harlaftis \& Greiner 2004). The accretion onto the black hole yields highly variable X-ray emission (e.g., Belloni et al.\ 2000, and references therein). Still, even the hardest observed spectra are relatively soft, consisting of a blackbody-like component and a high-energy tail (e.g., Vilhu et al.\ 2001). They are softer than those of other black-hole binaries in the hard state, which $EF_E$ spectra peak at $\sim 100$ keV (e.g., Cyg X-1, Gierli\'nski et al.\ 1997), but generally similar to their intermediate and soft states (e.g., XTE J1550--564, Gierli\'nski \& Done 2003; Cyg X-1, Gierli\'nski et al.\ 1999; LMC X-1, LMC X-3, Wilms et al.\ 2001). This is confirmed by the comprehensive analysis of the states of \source\ by Done, Wardzi\'nski \& Gierli\'nski (2004), who attribute the lack of the hard state in \source\ to its very high, Eddington or higher, luminosity. 

The blackbody component arises, most likely, in an optically-thick accretion
disc. On the other hand, there were some uncertainty regarding the origin
of the tail. Three main models proposed involved Comptonization of
the blackbody photons by high-energy electrons. They differed, however, in the
distribution (and location) of the electrons, which are assumed to be either
thermal (Maxwellian), non-thermal (close to a power law), or in a free fall
onto the black hole.

A discussion of these models is given in Zdziarski (2000), who shows that the
thermal and free-fall models of the soft state of black hole binaries can be
ruled out, mostly by the marked absence of a high-energy cutoff around 100 keV
in the \gro\/ data (Grove et al.\ 1998; Gierli\'nski et al.\ 1999; Tomsick et
al.\ 1999; Zdziarski et al.\ 2001, hereafter Paper I; McConnell et al.\ 2002). In particular, those two models have been shown to be strongly ruled out for \source\ in Paper I. The present best soft-state model involves electron acceleration out of a Maxwellian distribution (i.e., a non-thermal process), which leads to a hybrid electron distribution consisting of both thermal and non-thermal parts (Zdziarski, Lightman \& Macio{\l}ek-Nied\'zwiecki 1993; Poutanen \& Coppi 1998; Gierli\'nski et al.\ 1999; Coppi 1999).

In order to gain further insight into the nature of radiative processes in
\source, we study here available broad-band, $\sim$2--$10^3$ keV, spectra
and variability of this source from simultaneous observations by the PCA and HEXTE detectors on board \xte\/ and the OSSE detector on board \gro. An earlier study of two of those spectra corresponding to the lowest and highest X-ray flux was given in Paper I. The spectra, showing extended power laws without any cutoff up to at least 600 keV, provide strong evidence for the presence of non-thermal Comptonization. Here, we also study the rms spectra from the PCA data and their implications for physics of the source. Given the importance of the distance and inclination of the system for our spectral fits and their theoretical interpretation, we begin our analysis with estimating those parameters in Section \ref{distance} below. 

\section{The distance and inclination}
\label{distance}

Using the rotational velocity of the donor, $v \sin i =26 \pm 3$ km s$^{-1}$ (Harlaftis \& Greiner 2004), and assuming synchronous rotation with the orbital period, $P_{\rm orb} = 33.5$ d, we find the donor radius of $R_2 = (17.2 \pm 2.0)\sin^{-1} i\, {\rm R}_{\sun}$. The reddening-corrected secondary magnitude is $K_0=12.1$--12.6, as obtained using the unveiled magnitude of the donor, $K = 14.5$--15 (Greiner et al.\ 2001b), and correcting it for the extinction of $A_{K}= 2.4 \pm 0.2$ (Chapuis \& Corbel 2004). Then, we use the Barnes-Evans relation (Cahn 1980),
\begin{equation}
F_K = 4.2211 - 0.1 K - 0.5 \log_{10} s,
\end{equation}
where the $K$ surface brightness is $F_K = 3.84$ ( Beuermann, Baraffe \& Hauschildt 1999), which yields the angular diameter of the secondary of $s=0.0220$--0.0175 mas. This corresponds to the distance of 
\begin{equation}
d = (6.4\!-\!\!10.2)\sin^{-1} \!i\,  \mathrm{kpc},
\label{range_d}
\end{equation}
or $d=6.9$--11.3 kpc for $i=66\pm 2\degr$ (Fender et al.\ 1999; see also Mirabel \& Rodriguez 1994).

Other estimates include $6 \la d  \la 12$ kpc from the kinematics of interstellar clouds (Dhawan Goss \& Rodriguez 2000a; Chapuis \& Corbel 2004), $d \leq 11.2 \pm 0.8$ kpc from radio data (Fender et al.\ 1999), $d \sim 12$ kpc from the secular (transverse) motion (Dhawan, Mirabel \& Rodriguez 2000b), and, the most accurate, distance of $d=12.1 \pm 0.8$ kpc implied by the radial systemic velocity (Greiner et al.\ 2001a). The mutual agreement of the last two estimates is consistent with the motion of the source being primarily due to the Galactic rotation. 

Recently, Kaiser et al.\ (2004) have identified two IRAS sources, IRAS 19124+1106 and IRAS 19132+1035, each located at the angular distance of $\sim 26'$ on opposite sides of the direction to \source, with the impact sites of the jets of that source. This identification resulted in a new distance estimate to \source, $d\simeq 6.5$ kpc, which, in combination the observed jet motions on small scales, yielded a jet velocity of about $0.7 c$ and an angle of 47--$53\degr$ of the jets to our line of sight. However, the two IRAS regions do not move together with \source. The radial velocities of the two IRAS sources are redshifted by about 70 km s$^{-1}$ (IRAS 19132+1035) and 60 km s$^{-1}$ (IRAS 19124+1106), respectively, with respect to the radial systemic velocity of \source\ of $-3 \pm 10$ km s$^{-1}$ derived from studies of $^{12}\rm CO$ and  $^{13}\rm CO$ lines in its K-band spectra (Greiner et al.\ 2001a). Kaiser et al.\ (2004) doubted the absolute wavelength calibration of the observed spectra and suggested that `systematic effects may well have altered the determination of the systemic velocity'; however they neither clarified nor proposed any such effects. On the other hand, it is known  from studies of symbiotic red giants that the radial velocity curves based on near-IR CO lines are very accurate, and any systematic effects are $<1$--2 km s$^{-1}$ (e.g., Fekel et al.\ 2000). In particular, there is an excellent agreement between the systemic velocity derived from CO and  optical lines. Although in the case of \source, the CO band accuracy is less than in typical sources due to relatively high veiling of the giant spectrum, it is unlikely that the error exceeds about 8 km s$^{-1}$ (J. Greiner, personal communication).  Moreover, this lower distance is incompatible with the lower limit derived from the red giant radius and brightness, which for the system inclination (= jet inclination) of $i =53\degr$ yields, using equation (\ref{range_d}), $d=8.0$--12.8 kpc, i.e., substantially more than the distance used by Kaiser et al.\ (2004) to obtain that inclination.

Thus, the two IRAS sources appear not to be associated with \source. Combining other constraints above, we find $d\simeq 11$ kpc as the only value approximately compatible with all of them. Hereafter we assume this distance and $i=66\degr$.

\section{The data}
\label{data}

We use \xte\/ data simultaneous with those by the OSSE and showing relatively weak X-ray variability, which allows a meaningful analysis of the average spectra. In Paper I, the variability states used for spectral fitting were called $\chi$ and $\gamma$, according to the classification of Belloni et al.\ (2000). However, Klein-Wolt et al.\ (2002) introduced then an additional variability state, $\omega$, and Naik, Rao \& Chakrabarti (2002) identified the $\gamma$-state observation of Paper I as belonging to the state $\omega$, which name we use hereafter. 

We extract spectra from pointed \xte\/ observations using {\sc ftools} v.\ 
5.3. We add systematic errors of 1 per cent to each PCA and HEXTE channel. We extract spectra from both HEXTE clusters separately. Since those data sets were obtained with various number of PCA detector units (PCUs) operating, we have decided to use only the PCU 0 and 2, which were switched on in all of our data. We use only layer 1 of the PCA. In order to test the effect of changing the method of data extraction, we have also obtained the PCA spectra for all operating detectors for a given data set, and used all layers. However, we have found only negligible differences in the results of spectral fits to those data with respect to the PCU 0, 2, layer 1, configuration. We have also tested for the effect of the choice of a particular \xte\/ observation in cases when more than one exist (for the $\chi$ state, see Paper I), which effect has been found to be minor. Table \ref{tab:log} shows the log of the \xte\/ and OSSE data used.

\begin{table*}
\centering \caption{Log of the OSSE and \xte\/ observations.}
\begin{tabular}{cccccccc}
\hline
\multicolumn{3}{c}{OSSE} & \multicolumn{4}{c}{\it RXTE} \\
VP & Start & End & Observation ID & Date & Time (UT) & Variability state\\
\hline
601 & 1996-10-15 & 1996-10-29 & 10408-01-45-00 & 1996-10-29 & 11:56--17:30 & $\chi$ \\
619 & 1997-05-14 & 1997-05-20 & 20187-02-02-00 & 1997-05-15 & 11:31--18:42 & $\chi$ \\
720 & 1998-05-05 & 1998-05-15 & 30402-01-12-01 & 1998-05-12 & 00:28--01:00 & $\chi$ \\
813 & 1999-04-21 & 1999-04-27 & 40403-01-07-00 & 1999-04-23 & 01:57--02:44 & $\omega$ \\
917 & 2000-04-18 & 2000-04-25 & 50405-01-02-00 & 2000-04-22 & 09:21--10:19 & $\chi$ \\
\hline
\end{tabular}
\label{tab:log}
\end{table*}

\begin{figure}
\centerline{\psfig{file=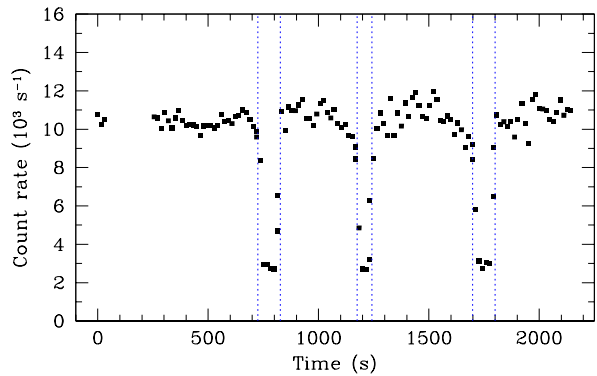,width=7.8cm}} \caption{The PCA
lightcurve of the observation in the $\omega$ state (VP 813), showing the
intervals used for calculating the high and low spectra. The origin of
the time axis corresponds to 1999-04-23 02:05:03 UT.
\label{813_lc} }
\end{figure}

Since we need to integrate over rather long periods in order to obtain accurate 
spectra, we have searched for flares/dips from the lightcurves prior to the 
spectral extraction. This affected the $\chi$-state \xte\/ spectrum of VP 619, 
were we removed a few flare-like events. Furthermore, the lightcurve of the 
$\omega$ state, VP 813, consists of steady emission regularly interrupted by 
short dips. For that spectrum, we have then obtained separately the \xte\/ 
spectra for the steady state and the dips, as shown in Fig.\ \ref{813_lc}. We note that the steady emission outside the dips is very similar to that of the class $\gamma$, which classification was used in Paper I. 

The OSSE detector accumulated spectra in a sequence of 2-min measurements of the 
source field alternated with 2-min, offset-pointed measurements of background.  
The background spectrum for each source field was derived bin-by-bin with a 
quadratic interpolation in time of the nearest background fields (Johnson et 
al.\ 1993). The OSSE data include energy-dependent systematic errors estimated 
from the uncertainties in the low-energy calibration and response of the 
detectors using both in-orbit and pre-launch calibration data. They range from 
0.03 at 50 keV to 0.003 at $\ga 150$ keV (where, however, statistical errors 
dominate). The OSSE data used here are the same as in Paper I.

Apart from the energy spectra, we also calculate the corresponding power density 
spectra (PDS) from PCA lightcurves with 1/128-s resolution. Each PDS is an 
average of several Fourier transforms calculated over 512-s intervals of the 
lightcurve. This gives the frequency band of 1/512--64 Hz. We create PDS both 
in the full 2--60 keV PCA energy band, and in three bands of approximately 2--7, 
7--15 and 15--60 keV.

In order to study energy dependence of variability in details, we also extract 
energy-dependent rms, using two different methods. In both methods, we first 
extract background-subtracted lightcurves with 1/128-s resolution in each PCA 
energy channel. In the first method, we calculate intrinsic or excess rms [see eq. (10) in Vaughan et al. 2003], characterizing the intrinsic source variability not related to the Poissonian noise. In the second method, we 
calculate the PDS from each of the lightcurves (over 512-s intervals), subtract the Poissonian noise, and correct for dead-time effects (Revnivtsev, Gilfanov \& 
Churazov 2000). The energy-dependent rms is then found by integrating the PDS over 1/512--64 Hz frequency band. Both methods give very similar results, see Section \ref{frac} below. 

\section{Spectral models}
\label{models}

We first fit the data with the hybrid Comptonization model, {\sc compps}, of Poutanen \& Svensson (1996), as implemented in XSPEC (Arnaud 1996). In this model, the temperature, $kT$, and the total Thomson optical depth, $\tau$, of the electrons are free parameters and the distribution is assumed to be Maxwellian up to the Lorentz factor, $\gmin$ (which is the third free parameter), and a power law with an index, $\Gamma$, above it up to $\gmax$, assumed here to be $10^3$. The iterative scattering method of Poutanen \& Svensson (1996) is not applicable to plasmas with large $\tau$, and we constrain $\tau$ to $\leq 6$.  We assume a spherical geometry of the hot plasma (the model geometry parameter is set to 0). Since some of the data may only weakly constrain the form of the nonthermal tail, we assume $\Gamma\geq 2$, which does not allow for very hard tails, unlikely to be produced by a physical acceleration process. The seed photons for Comptonization are assumed to have the spectrum of a blackbody disc (Mitsuda et al.\ 1984) with the maximum temperature of $kT_{\rm bb}$. 

We also take into account Compton reflection with a solid angle of $\Omega$
(Magdziarz \& Zdziarski 1995) and an Fe K$\alpha$ emission from an accretion
disc (Shakura \& Sunyaev 1973) assumed to extend down to $100GM/c^2$. We allow the reflecting surface to be ionized, with the ionization parameter of $\xi\equiv L_{\rm ion}/n R^2$, where $L_{\rm ion}$ is defined as the 5 eV--20 keV luminosity in a power law spectrum and $n$ is the density of the reflector located at distance $R$ from the illuminating source. We use here the ionization model by  Done et al.\ (1992), which is not applicable to very highly ionized Fe (Ballantyne, Ross \& Fabian 2001), but sufficiently accurate at low/moderate ionization, such as that found in our data. Consistent with the found low ionization, we keep the line energy fixed at 6.4 keV. The reflector temperature is kept at $10^6$ K. The elemental abundances of Anders \& Ebihara (1982), an absorbing column of $\nh\geq \nh^{\rm G}$, and $i=66\degr$ (Section \ref{distance}) are assumed. Here, $\nh^{\rm G}$ is the Galactic column density in the source direction, which we take here as $\simeq 1.8\times  10^{22}$ cm$^{-2}$, following Dickey \& Lockman (1990). (We note, however, that a higher value of $\nh^{\rm G}$ has been claimed by Chaty et al.\ 1996.)

We then fit the data with the  model {\sc eqpair} (Coppi 1999; Gierli\'nski et 
al.\ 1999), which calculates self-consistently microscopic processes in a hot 
plasma with electron acceleration at a power law rate with an index, 
$\Gamma_{\rm inj}$, in a background thermal plasma with a Thomson optical depth 
of ionization electrons, $\tau_{\rm i}$. The  electron temperature, $kT$, is 
calculated from the balance of Compton and  Coulomb energy exchange, as well as 
\ee\ pair production (yielding the total optical depth of $\tau>\tau_{\rm i}$) 
is taken into account.  The last two processes depend on the plasma compactness, 
$\ell\equiv {\cal  L}\sigma_{\rm T}/({\cal R} m_{\rm e} c^3)$, where ${\cal L}$ 
is a power  supplied to the hot plasma, ${\cal R}$ is its characteristic size, 
and  $\sigma_{\rm T}$ is the Thomson cross section. We then define a hard 
compactness, $\lh$, corresponding to the power supplied to the electrons, and a 
soft compactness, $\ls$, corresponding to the power in seed photons (of a disc blackbody spectrum) irradiating the plasma. We assume $\ls=100$, which is 
consistent with the source luminosity and estimates on the size of the hot 
plasma, see Paper I. The compactnesses corresponding  to the electron 
acceleration and to a direct heating (i.e., in addition to Coulomb energy 
exchange with non-thermal \ee\ and Compton heating) of the  thermal \ee\ are 
denoted as $\lnth$ and $\lth$, respectively, and $\lh=\lnth +  \lth$. Details of 
the model are given in Gierli\'nski et al.\ (1999).

The nonthermal electrons are accelerated between the Lorentz factors $\gmin$ and $\gmax$, with the latter assumed to be $10^3$ (as in the {\sc compps} model).
For $\gmin$, we use the best-fit value of the {\sc compps}, as $\gmin$ in {\sc eqpair} only weakly affects the fit results and thus cannot be reliably fitted. Reflection and absorption is treated the same way as in {\sc compps}. 

We note that results from the \gro/EGRET, which did not detect GRS 1915+105, yield an upper limit at $\sim$100 MeV of $EF_E\la 0.04$ keV cm$^{-2}$ s$^{-1}$ (Hartman et al.\ 1999). This upper limit is fully satisfied in all of {\sc eqpair} fits to the data obtained by us below, since that model takes into account photon absorption in pair-producing photon-photon collisions. Also, that model treats self-consistently energetic electrons and positrons produced in such events. On the other hand, the {\sc compps} model does not include pair production, and some of our fit results may violate the EGRET upper limit. Thus, our results obtained with {\sc compps} should be treated as applicable only to the photon energy range up to soft \g-rays, and our preferred overall model is {\sc eqpair}. On the other hand, {\sc compps} treats Comptonization more accurately as well as it gives us directly the steady-state electron distribution in the source. 

The uncertainties ranges for each parameter below are given for 90 per cent confidence, i.e., $\Delta \chi^2=2.71$. To assess the significance of the nonthermal electrons, we also fit the data with purely thermal models (i.e., without any high-energy tail), using both {\sc compps} and {\sc eqpair}. We then use the F-test (Bevington \& Robinson 1992) to calculate the probability, $P_{\rm th}$, that the fit improvement by allowing a nonthermal electron tail (with two additional free parameters, the index and normalization) is by chance. 

\begin{figure}
\centerline{\psfig{file=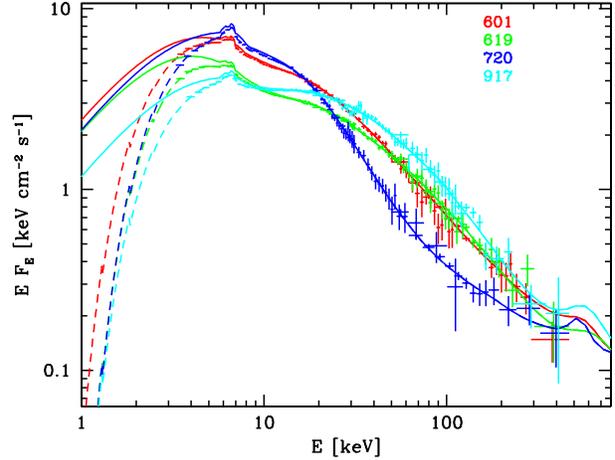,width=8cm}}
\caption{Broad-band spectra in the $\chi$ state. The solid and dashed curves give the intrinsic and absorbed {\sc eqpair} model spectra, respectively. These and subsequent spectra are shown normalized to the PCA data.
\label{chi_eeuf} }
\end{figure}

\section{Broad-band spectra and PCA power spectra}
\label{spectra}

The fit results for both {\sc compps} and {\sc eqpair} models are given in Table 2. We find rather good agreement between those results, in particular for the blackbody and electron temperatures and the absorbing column and the characteristics of reflection. The values of $\tau$ are somewhat different as they are defined in different way in the two models. Note that in most cases $\Gamma> \Gamma_{\rm inj}$, due to the softening of the steady-state electron spectrum (fitted in {\sc compps}) with respect to the spectrum of the injected electrons (fitted in {\sc eqpair}). Also, the results for VP 619 and 813 are similar to those given in Paper I. In all cases, the values of the relative normalization between the PCA, HEXTE and OSSE spectra are very close to unity. 
The values of $\cnu<1$ are due to the systematic errors being apparently somewhat overestimated. 

The bolometric isotropic fluxes given in Table 2 correspond to $L/\ledd\approx 7\times 10^6 F_{\rm bol} (D/11\,{\rm kpc})^2 (14\msun/M)$, where $\ledd\approx 1.5(M/\msun)\times 10^{38}$ erg s$^{-1}$ is the Eddington luminosity. They yield $L/\ledd\approx 0.17$--0.24 and 0.5--0.6 for the $\chi$ states and the high-flux part of the $\omega$ state, respectively. 

We confirm the conclusion of Paper I that the commonly-used model of a disc blackbody and an e-folded power law provide very bad fits to all of our data, with $\cnu \gg 1$. On the other hand, all our fits with the physical Comptonization model yield $\cnu< 1$, see Table 2. All the data rule out purely thermal Comptonization, with extremely low values of the corresponding probability, $P_{\rm th}$. Below, we discuss our results separately for the $\chi$ and $\omega$ states. 

\subsection{The $\chi$ state}
\label{chi}

\begin{figure}
\centerline{\psfig{file=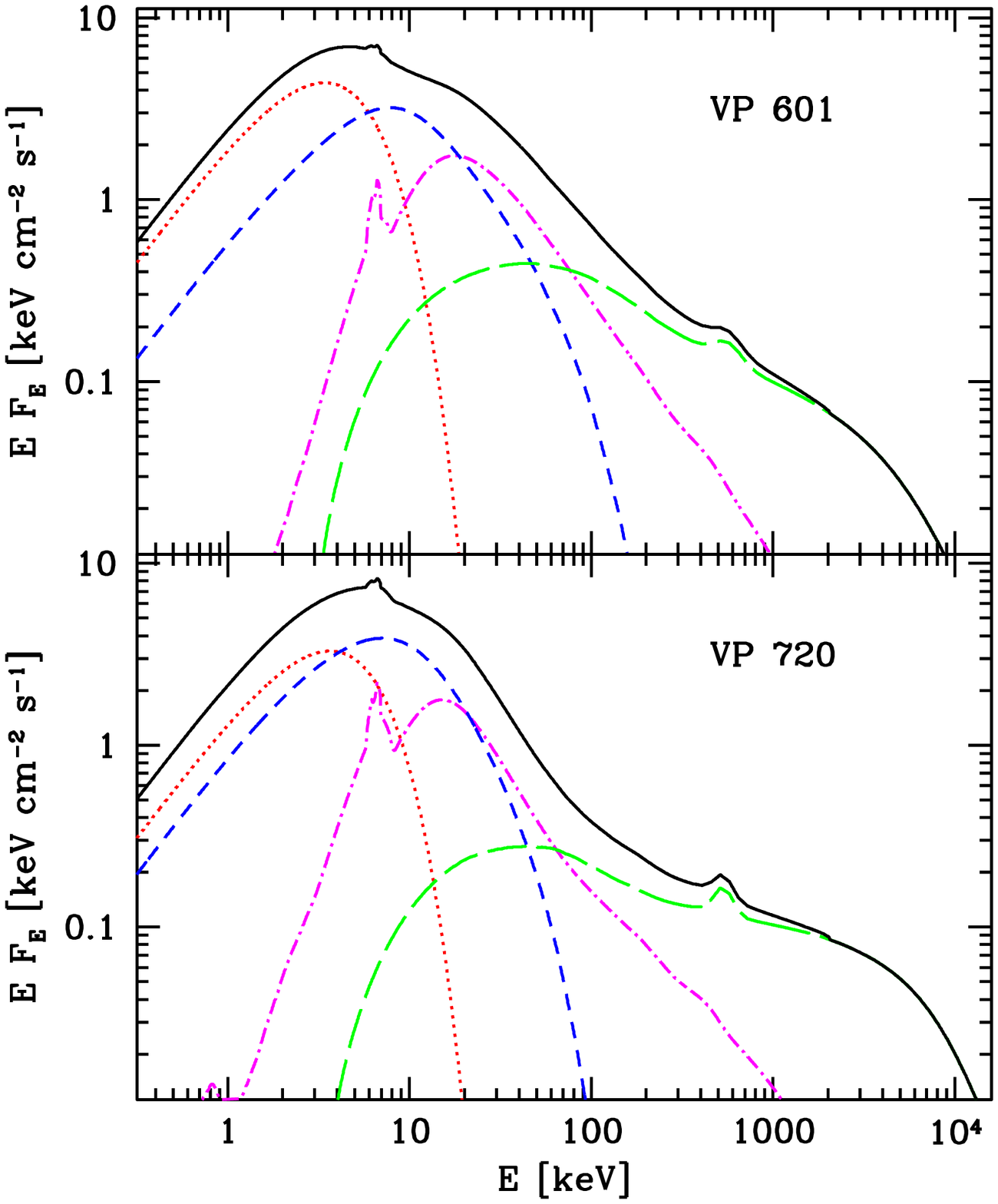,width=7.3cm}}
\centerline{\psfig{file=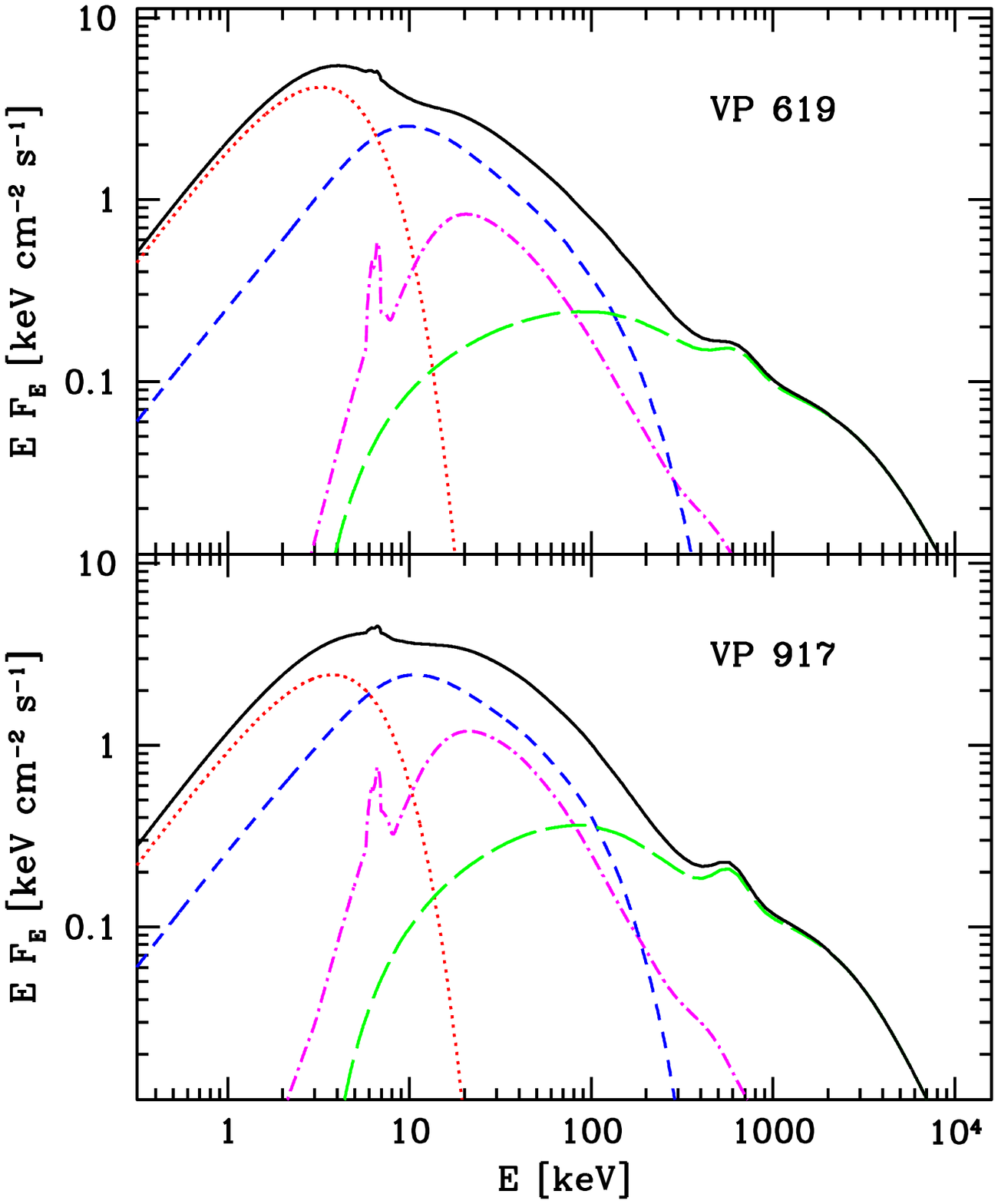,width=7.3cm}}
\caption{Spectral components of the {\sc eqpair} fits (solid curves) to the $\chi$ state. The dots show the unscattered blackbody, the short and long dashes
show Compton scattering by thermal and nonthermal electrons, respectively (the latter also includes the pair annihilation component), and the dot-dashed curves show Compton reflection including the Fe K$\alpha$ line. 
\label{chi_models} }
\end{figure}

Fig.\ \ref{chi_eeuf} shows the unfolded spectra and the best-fit {\sc eqpair} models for our four data sets in the $\chi$ state. We see all the spectra are rather similar, with relatively modest variability on timescales between the observations. Two pairs of spectra differ by the level of the $\sim$10 keV flux (high: VP 601, 720, low: VP 619, 917) but there is no obvious correlation with the level of the high-energy, $\sim$100 keV flux. 

\begin{figure}
\centerline{\psfig{file=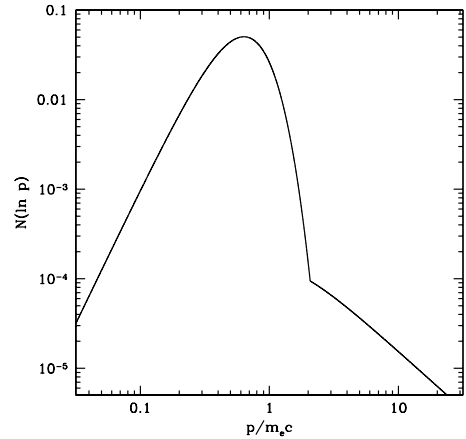,width=6cm}}
\caption{The distribution of the Comptonizing electrons corresponding to the best fit with the {\sc compps} model to the $\chi$-state data of the VP 619. See Table 2 for the parameters. The distribution of the momentum, $N(\ln p)=p N(p)$ is shown, and its normalization is arbitrary. The distibution is Maxwellian up to the $p$ corresponding to the $\gmin$ and a power law, $\propto \gamma^{-\Gamma}$ above it.
\label{electron} }
\end{figure}

The two spectra with the high 10-keV flux also show relatively strong reflection, see Table 2. The spectral components for the four spectra are shown in Fig.\ \ref{chi_models}. Fig.\ \ref{electron} shows the electron distribution corresponding to the {\sc compps} best-fit model to the spectrum of the VP 619.

The thermal Compton component shown in Fig.\ \ref{chi_models} corresponds 
to emission from purely thermal plasma of the same $\tau$ and $kT$ (adjusted in {\sc eqpair} by a suitable reduction of $\lh/\ls$) as the thermalized electrons in the hybrid distribution. The non-thermal component is the difference  between the total scattering spectrum by the hybrid electrons and the thermal-scattering spectrum. See Hannikainen et al.\ (2005) for details of the method. 

We see in Fig.\ \ref{chi_models} that our models predict the power-law emission extending with no cutoff up to several MeV and weak annihilation feature (with the plasma being pair-dominated in some cases, e.g., with $\tau_{\rm i} \rightarrow 0$ allowed for VP 619 and 917). These predictions should be tested by future soft \g-ray missions. On the other hand, we predict the emission at $\sim$100 MeV to be very weak due to pair absorption, and below the detectability level for {\it GLAST\/} (unlike the case of Cyg X-1, see Zdziarski \& Gierli\'nski 2004). 

\begin{figure}
\centerline{\psfig{file=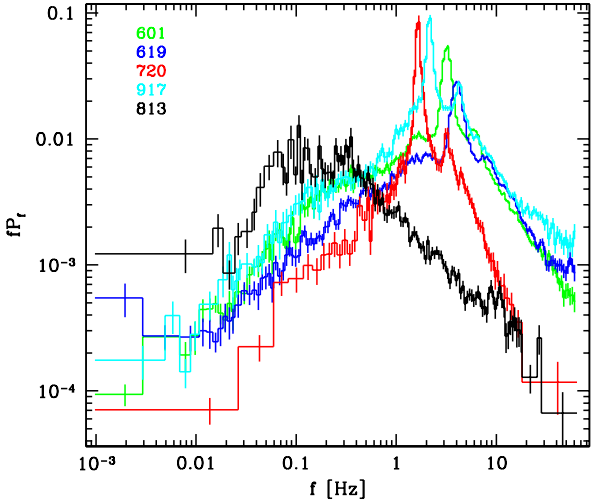,width=7.8cm}}
\caption{Comparison of the $\sim$2--60 keV power spectra for both the $\chi$ and $\omega$ states.
\label{power} }
\end{figure}

Fig.\ \ref{power} shows a comparison of all the power spectra (including that of the $\omega$ state, see Section \ref{chi}) for the entire PCA energy range, $\sim$2--60 keV. All the $\chi$-state power spectra look rather similar, with no obvious dependence on the fitted strength of reflection. The power spectrum of VP 720 is clearly the narrowest and shows the strongest QPO at the lowest frequency. The strong QPOs at frequencies of $\sim$1.5--4 Hz are characteristic to the intermediate/very-high state of black-hole binaries  (e.g., Zdziarski \& Gierli\'nski 2004). Also, the photon spectra (Figs.\ \ref{chi_eeuf}--\ref{chi_models}) and $L/\ledd\sim 0.2$ appear characteristic to those of the intermediate/very-high state (Zdziarski \& Gierli\'nski 2004) This confirms this classification of that state of GRS 1915+105 (Done et al.\ 2004) rather than as a hard/low state (Belloni et al.\ 2000). 

\begin{figure*}
\centerline{\psfig{file=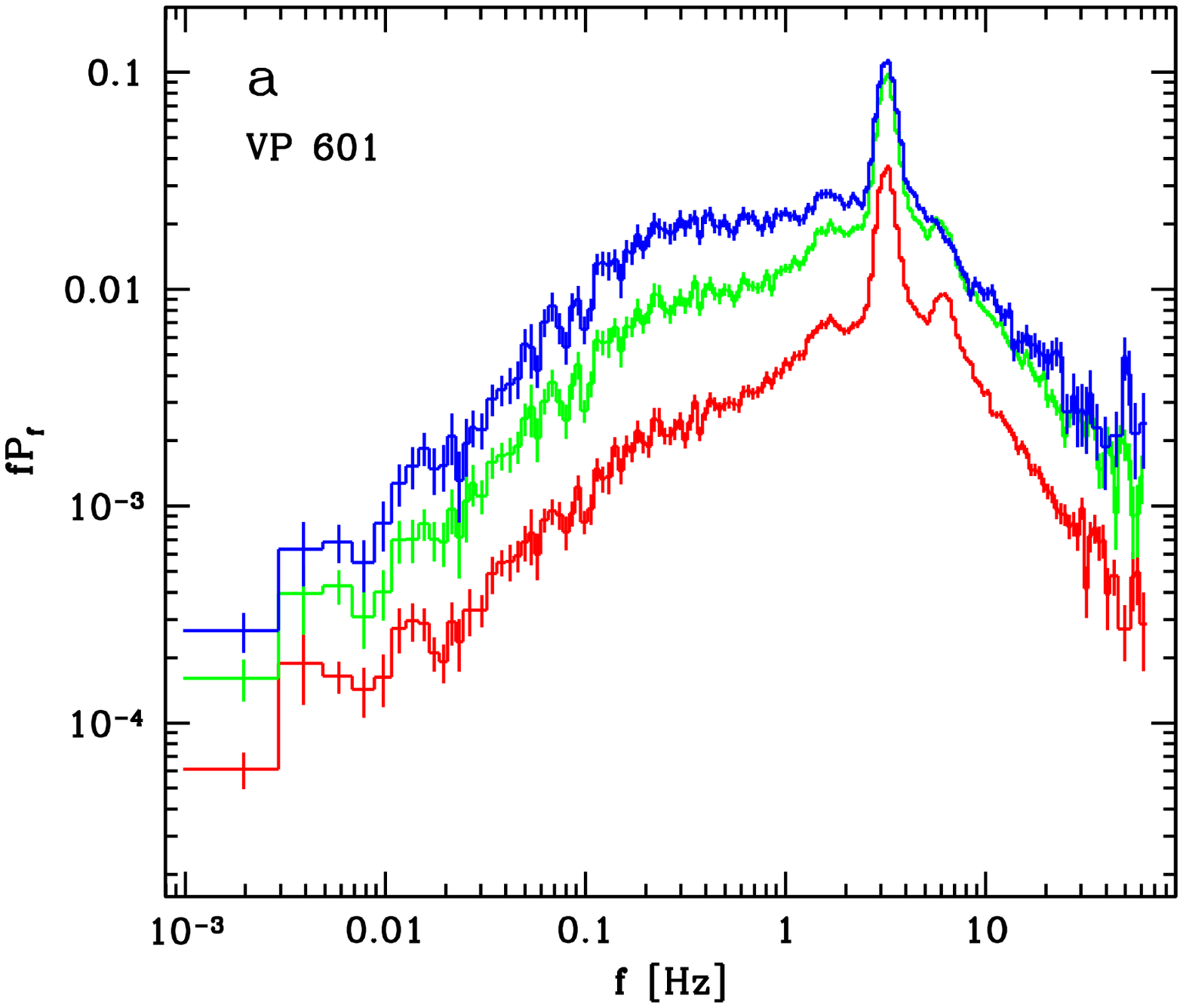,width=7.6cm}
\psfig{file=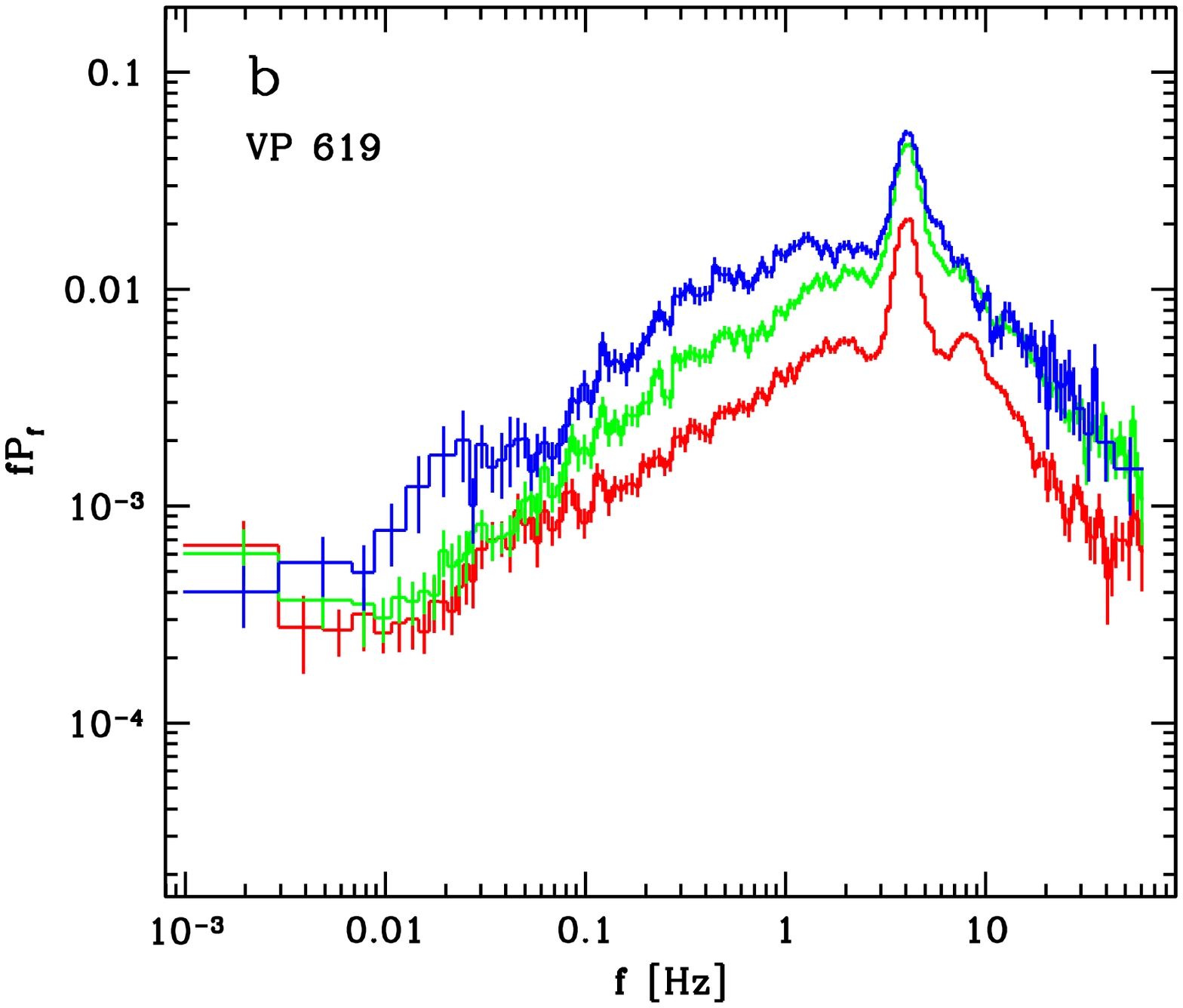,width=7.6cm}}
\centerline{\psfig{file=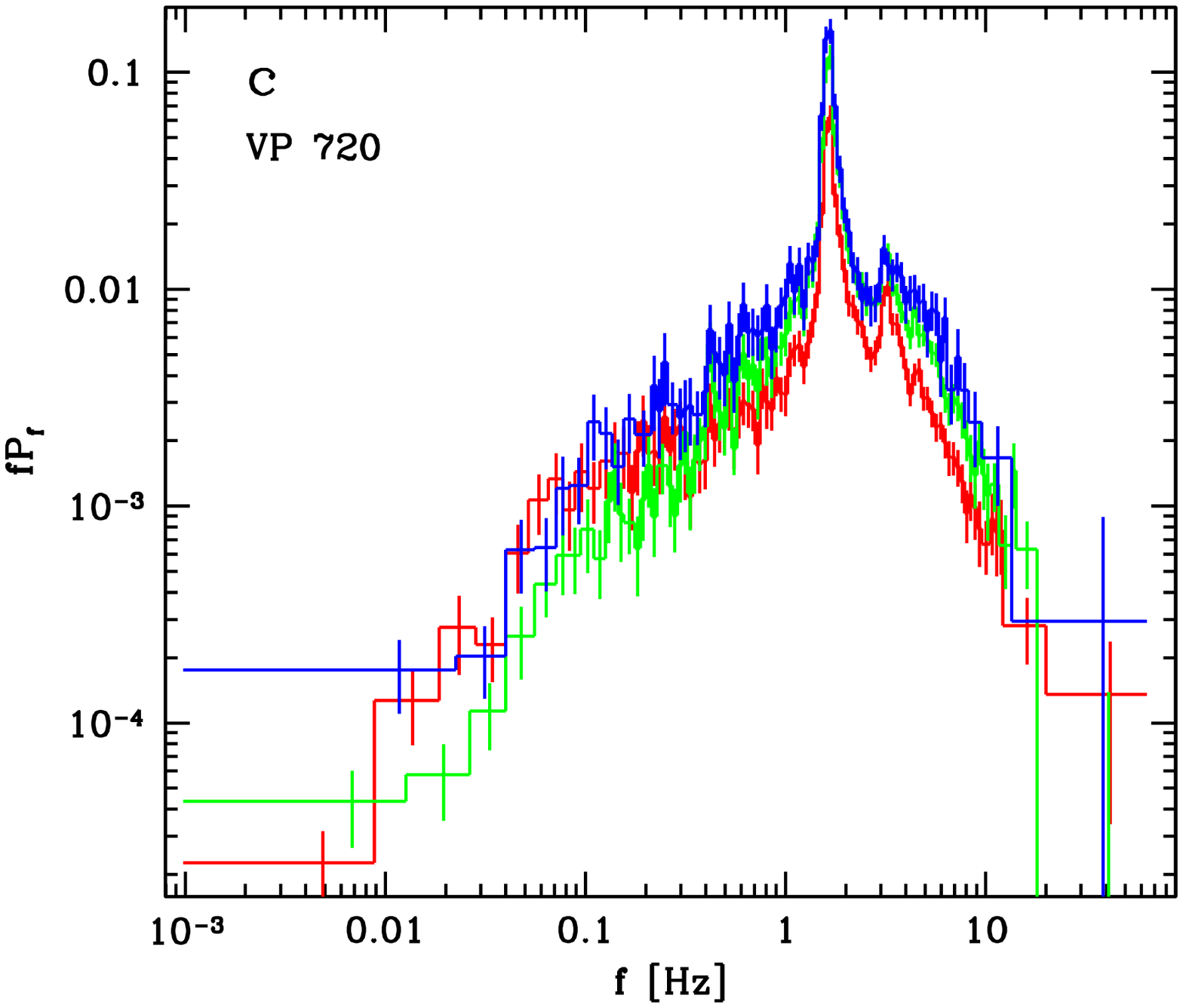,width=7.6cm}
\psfig{file=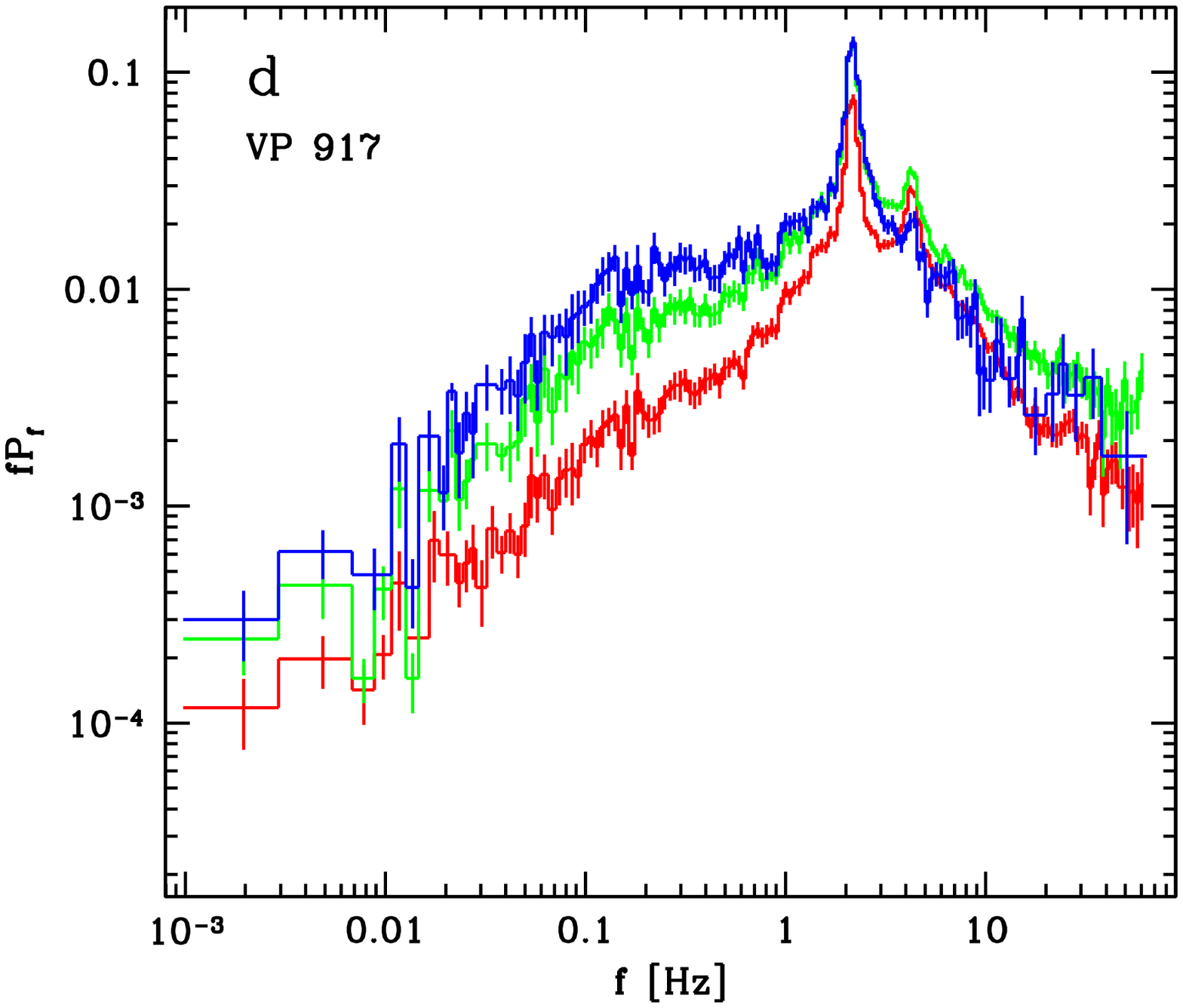,width=7.6cm}}
\caption{The power spectra of the individual data sets corresponding to the energy ranges of $\sim$2--7 keV (red), 7--15 keV (green) and 15--60 keV (blue).
\label{e_power} }
\end{figure*}

Fig.\ \ref{e_power} shows the photon energy-dependent power spectra for three energy channels. We see that the power generally increases with the photon energy. We also see that the width of the power spectra increases with the increasing photon energy. The main QPOs are surrounded by shoulders, showing weaker maxima at frequencies about a half and twice that of the QPO, with the low-frequency shoulders much more pronounced in the cases of VP 601 and 619. 

\subsection{The $\omega$ state}
\label{omega}

\begin{figure}
\centerline{\psfig{file=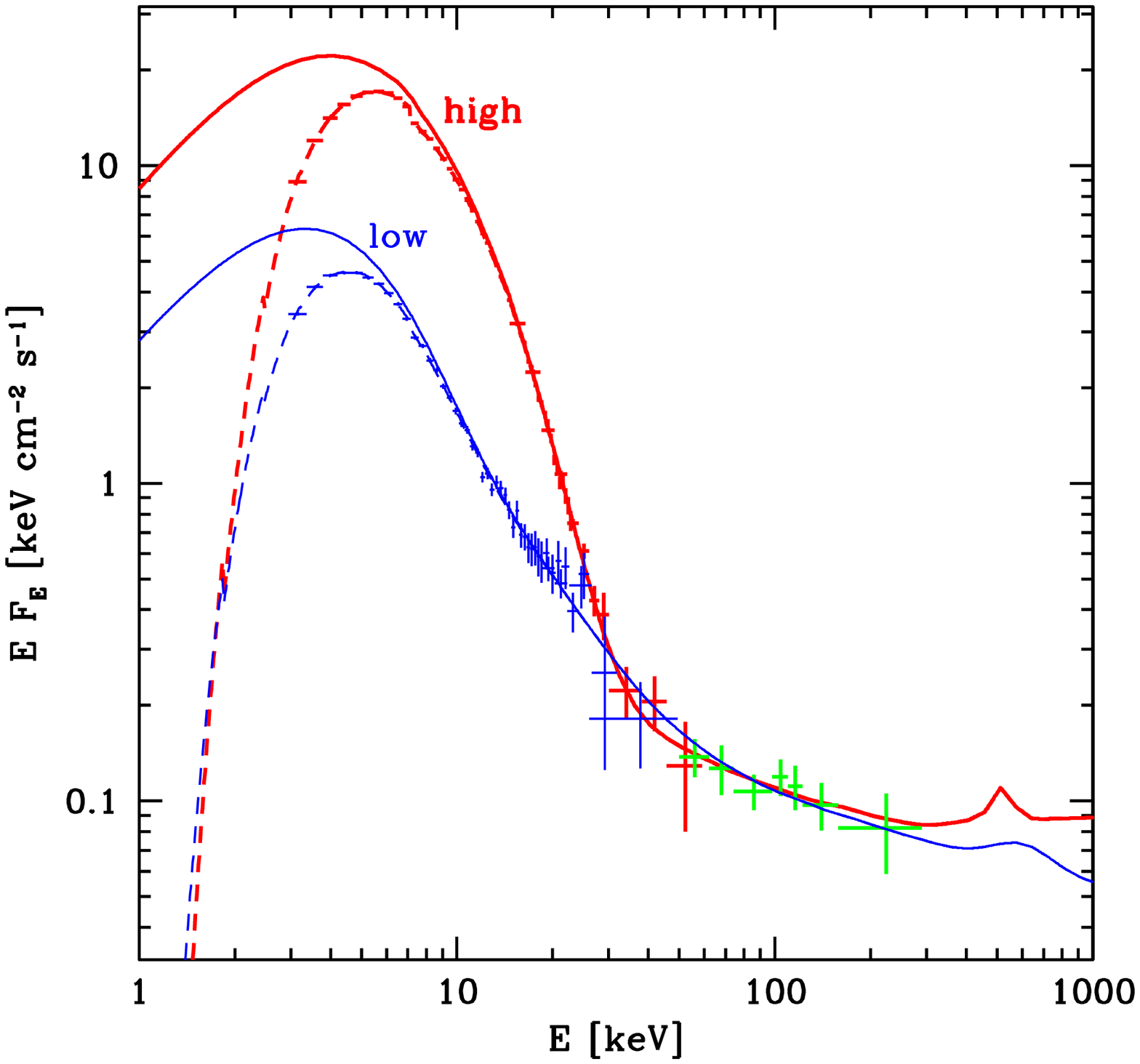,width=7.5cm}}
\caption{Broad-band spectra in the $\omega$ state. The heavy (red) and and light (blue) curves give the spectra corresponding to the high and low flux states. The OSSE spectrum (green) is assumed to be at its own normalization at both states, which is consistent with the HEXTE spectra observed. The solid and dashed curves give the intrinsic and absorbed models, respectively. 
\label{813_eeuf} }
\end{figure}

Fig.\ \ref{813_eeuf} shows the unfolded spectra of the VP 813 $\omega$ state, corresponding to the high-level continuum and to the dips. The relative normalization of the OSSE spectrum is assumed to be unity (i.e., the same as of the PCA) for both broad-band spectra. This is due to having at our disposal only the integrated OSSE spectrum, with the relatively low statistics. The HEXTE data agree with the OSSE well. The two PCA/HEXTE spectra pivot at $\sim$50 keV, and are thus consistent with the same level of the high-energy tail. 

The earlier version of the VP-813 data analyzed in Paper I required neither Compton reflection nor Fe K$\alpha$ line. In contrast, we find that the present high-flux data do require reflection, with the characteristic pattern of residuals in the fit without reflection with an edge at $\sim$7 keV and a hump at $\ga$10 keV. The probability of reflection appearing by chance is $\sim$0.01, 0.004 for the {\sc compps} and {\sc eqpair} fits, respectively, using the F-test. Still, we cannot be sure it is not due either to some systematic effect in the data or to complexity of the X-ray absorber. Thus, we give in Table 2 the results both allowing reflection and kept at null. On the other hand, the low-flux data do not require the presence of reflection, though the upper limits are relatively large, $\Omega/2\upi<1.2$, 1.9 in the {\sc compps} and {\sc eqpair} models, respectively.

\begin{figure}
\centerline{\psfig{file=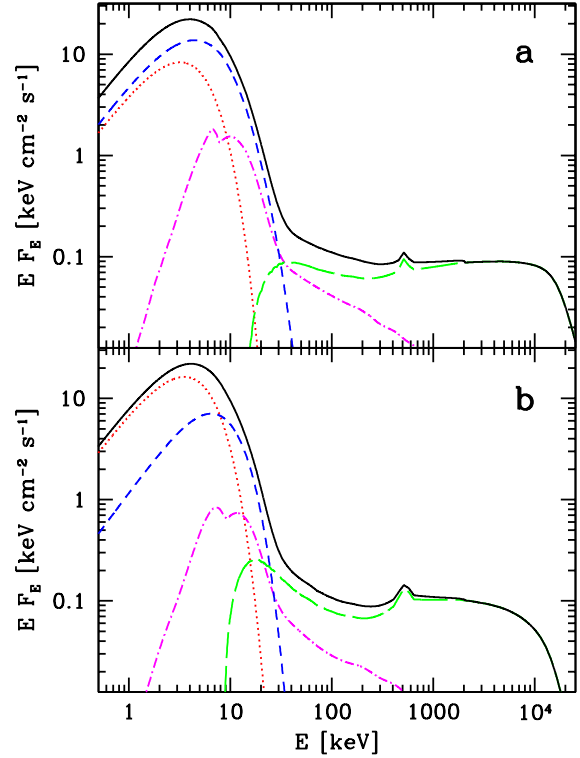,width=7.5cm}}
\caption{Spectral components in the $\omega$-state spectra (solid curves). The dots show the unscattered blackbody, the short and long dashes show Compton scattering by thermal and nonthermal electrons, respectively (the latter also includes the pair annihilation component), and the dot-dashed curves show Compton reflection. The panels (a) and (b) correspond to our main {\sc eqpair} model with reflection (Table 2) and to the model (DH) with an additional disc blackbody component, respectively, see Section \ref{omega}. In (b), the dotted curve corresponds to the additional disc blackbody only, and the unscattered part of the emission of the Comptonizing plasma is not shown.
\label{813_model} }
\end{figure}

Fig.\ \ref{813_model}(a) shows decomposition of the total intrinsic spectrum for the high-flux data as obtained by fitting the {\sc eqpair} model with reflection (Table 2). The high-flux spectrum and $L/\ledd\sim 0.5$ are roughly similar to ultrasoft spectra of black-hole binaries (e.g., Zdziarski \& Gierli\'nski 2004), although here strong Comptonization of the blackbody component, reflection, and relatively strong variability (Fig.\ \ref{power}, see below) are required, in contrast to the standard ultrasoft state (Gierli\'nski \& Done 2004). 

However, we have found that the decomposition of the prominent X-ray hump below $\sim$20 keV into the unscattered disc blackbody and its thermal Comptonization is not completely unique. We have considered a model with a separate (unscattered) disc blackbody component ({\sc diskbb}) in addition to one with a hybrid plasma Comptonizing disc blackbody photons ({\sc eqpair}) and allowing the two maximum blackbody temperatures of the direct disc emission and of the seed photons to the hybrid plasma to be different. This may correspond to a situation in which the plasma covers only an inner part of the disc while its outer part is seen directly. This model (hereafter abbreviated as DH) yields a number of relatively shallow local minima. In particular, we find a spectral solution with the maximum blackbody temperatures of the two components above of $1.48$ and $2.43$ keV, respectively (the latter value requires the color correction to be relatively large) at $\cnu=74/92$, similar to the corresponding value of 76/94 in Table 2. The main parameters of the hybrid plasma are $\lh/\ls=0.39$, $\lnth/\lh=0.76$, $\tau_{\rm i}=11.3$, $\Gamma_{\rm inj}=2.4$, $\Omega/2\upi =0.7$. The components of this model are shown in Fig.\ \ref{813_model}(b). This model also required relatively strong relativistic smearing of Compton reflection (consistent with the model geometry), and thus we have fixed the inner disc radius at $10GM/c^2$. 

Although the fits with the hybrid model to the broad-band spectra are very good, we cannot rule out the origin of the high-energy tail from a separate process. We have thus fitted the model (hereafter abbreviated as TP) consisting of scattering of the disc blackbody emission by thermal electrons only (using {\sc eqpair} with null reflection for the high-flux spectrum) and a separate power-law component. The obtained parameters for the thermal scattering of the disc blackbody are very similar to those given in Table 2, except that $\lh/\ls$ is somewhat lower, $0.27^{+0.02}_{-0.01}$ (since $\lh$ now does not include the nonthermal power). The fitted power law index is $\Gamma=2.1^{+0.2}_{-0.2}$, $kT_{\rm bb}=1.38^{+0.04}_{-0.03}$ keV, $\tau_{\rm i}=4.6^{+0.4}_{-0.1}$, and $\cnu=90/97$.

We note that the {\sc compps} fails to represent the high-flux spectrum because of the too high Thomson optical depth required by the data. The iterative scattering method (Poutanen \& Svensson 1996) used here utilizes only 50 scatterings, which number is much too low for the present case. The fitted value of $\tau$ becomes then artificially very high. In our model in Table 2, we constrain $\tau$ to 15, but this value is very strongly overestimated. Still, other parameters obtained using {\sc compps} (e.g., $kT$) appear correct, and they are in agreement with those obtained using {\sc eqpair}. 

\begin{figure}
\centerline{\psfig{file=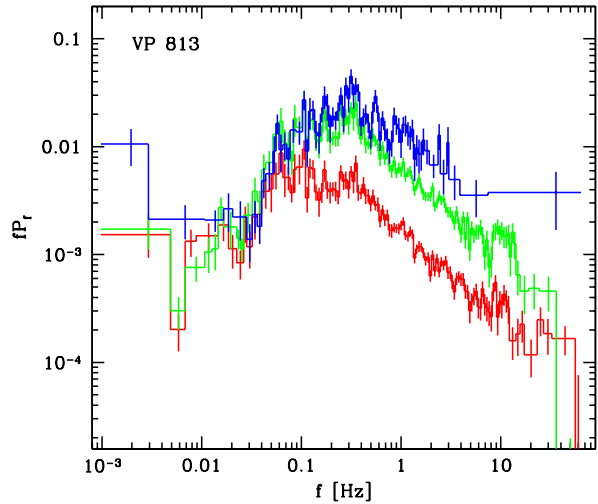,width=7.8cm}}
\caption{The power spectra of the $\omega$ state corresponding to the energy ranges of $\sim$2--7 keV (red), 7--15 keV (green) and 15--60 keV (blue).
\label{813_power} }
\end{figure}

Fig.\ \ref{power} shows the comparison of the total PCA power spectrum of the entire VP-813 observation to those in the $\chi$ state, and Fig.\ \ref{813_power} shows the energy-dependent PCA power spectra. We see the power spectra are distinctly different from those in the $\chi$ state. They have broad and flat (in power per logarithm of frequency) maxima at $\sim$0.05--0.4 Hz and are cut off at both ends.

\section{Fractional variability}
\label{frac}

\subsection{Data and multicomponent rms models}
\label{frac_m}

\begin{figure*}
\centerline{\psfig{file=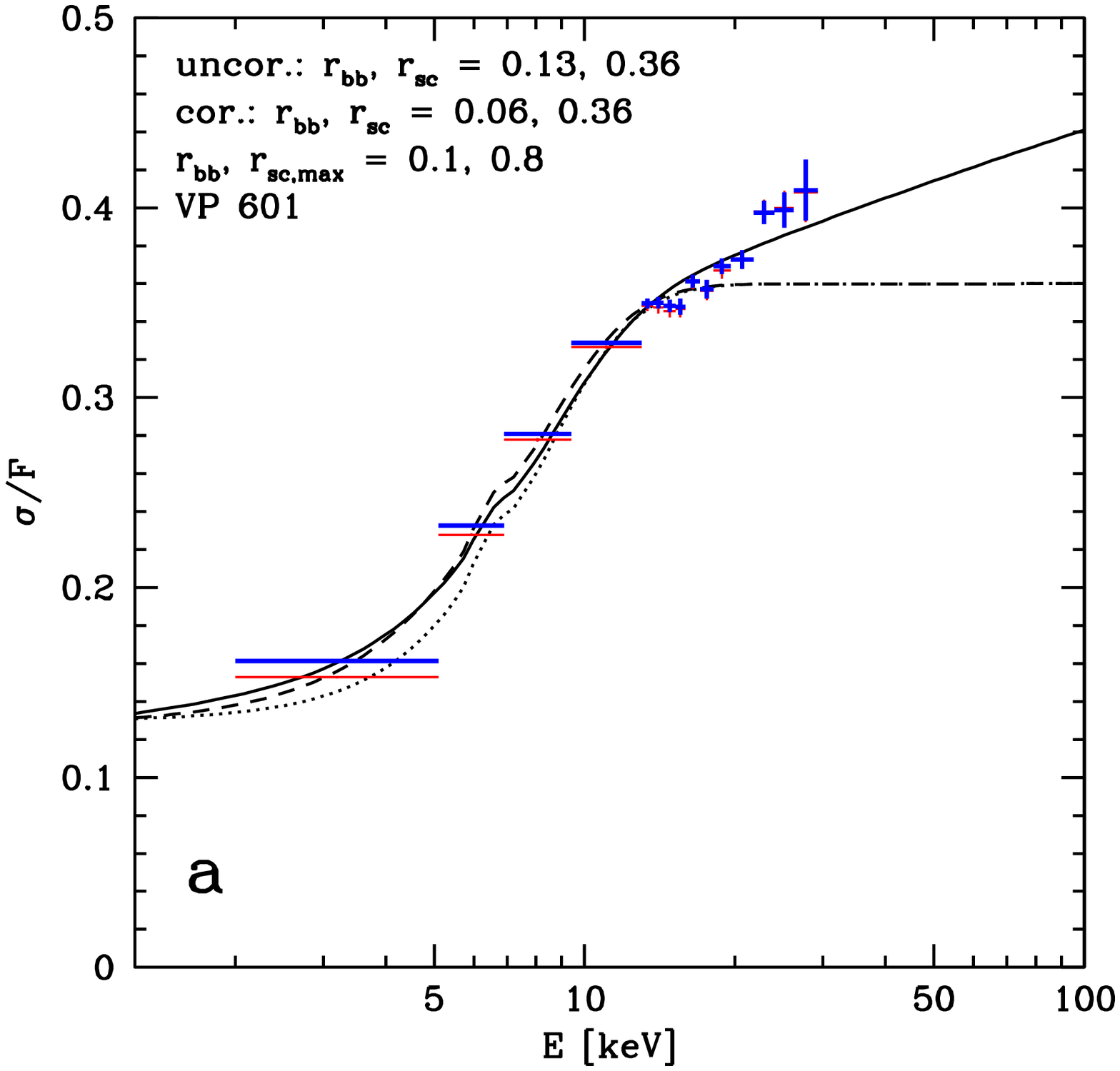,width=7.6cm}
\psfig{file=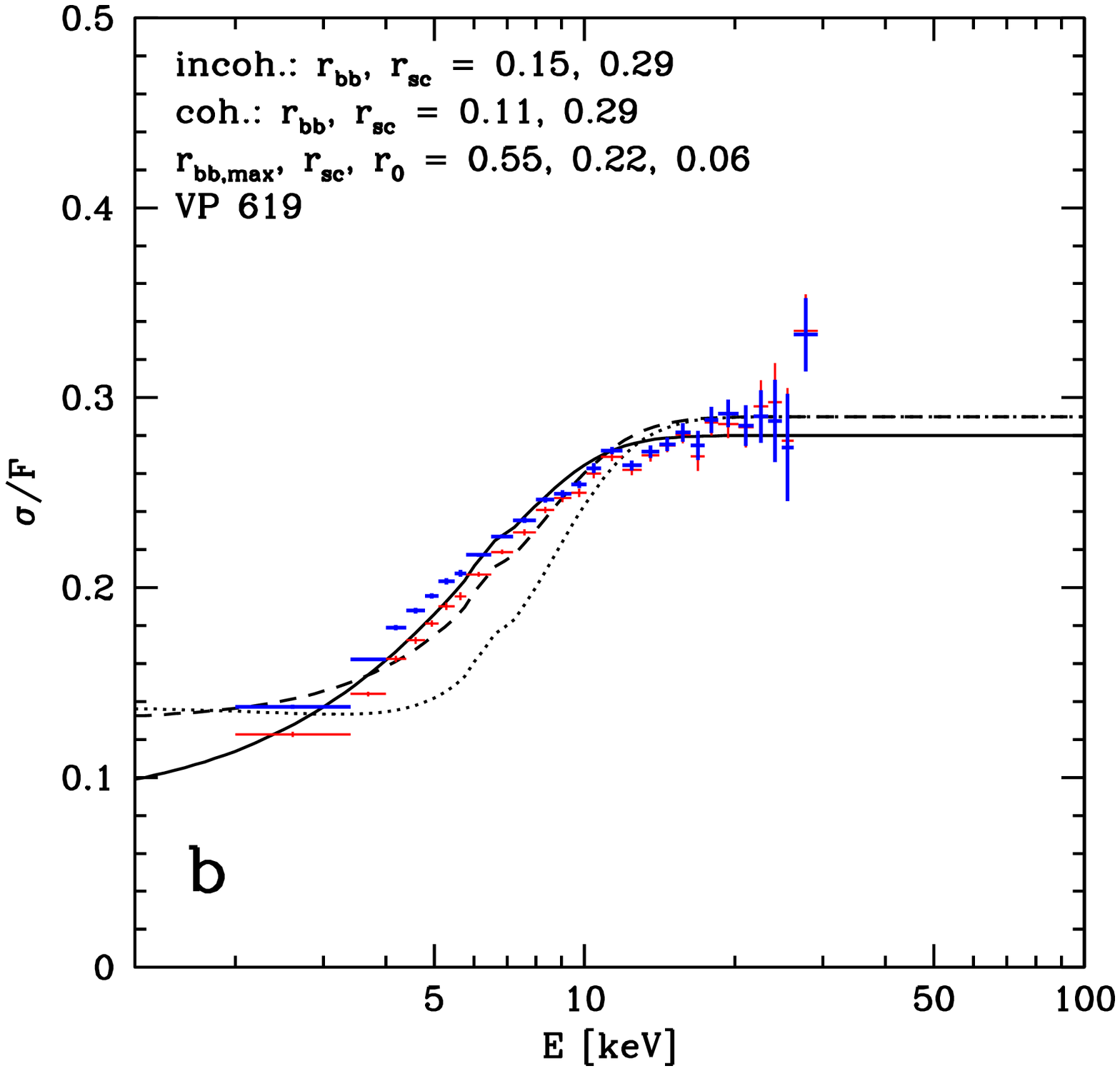,width=7.6cm}}
\centerline{\psfig{file=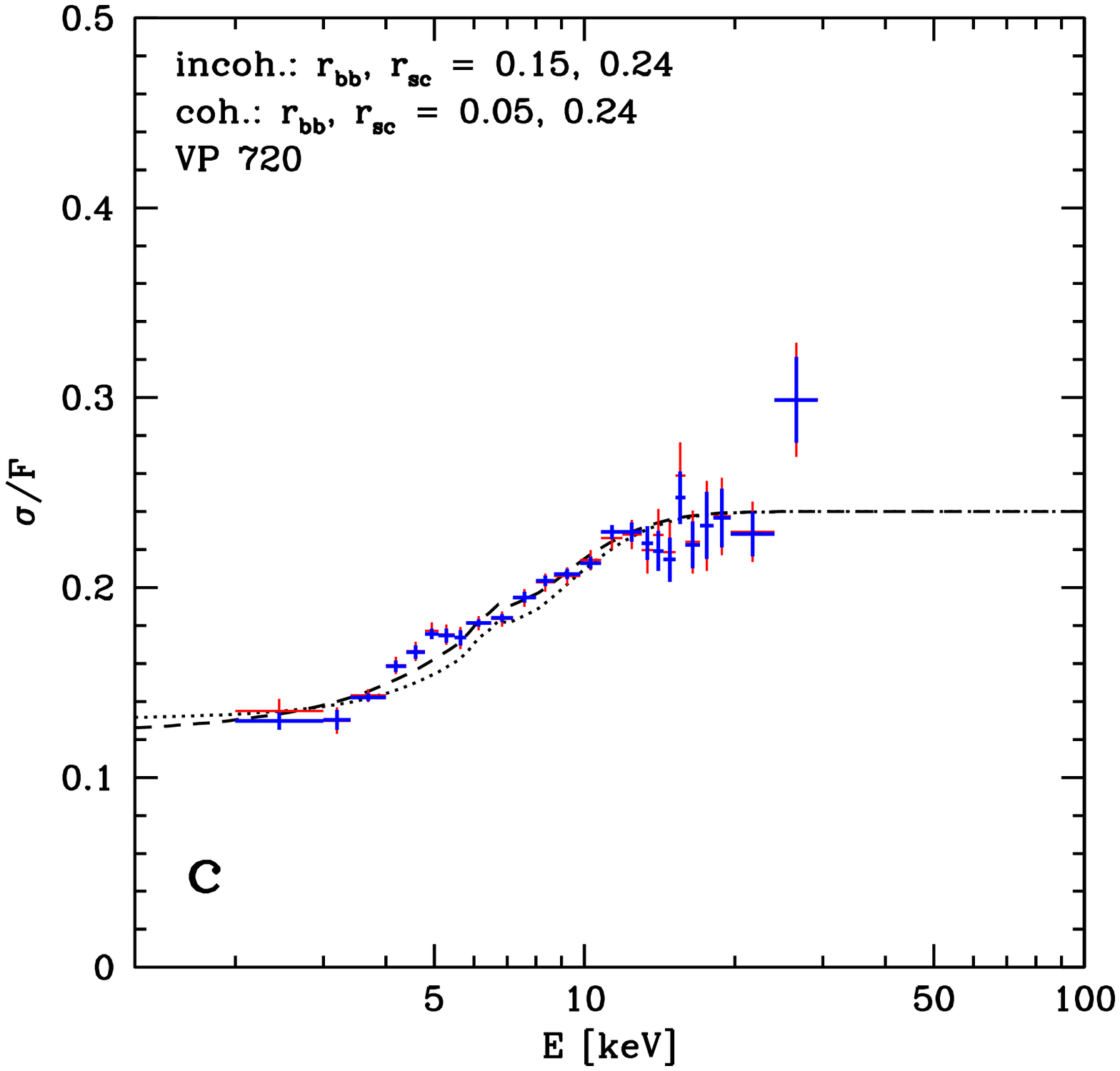,width=7.6cm}
\psfig{file=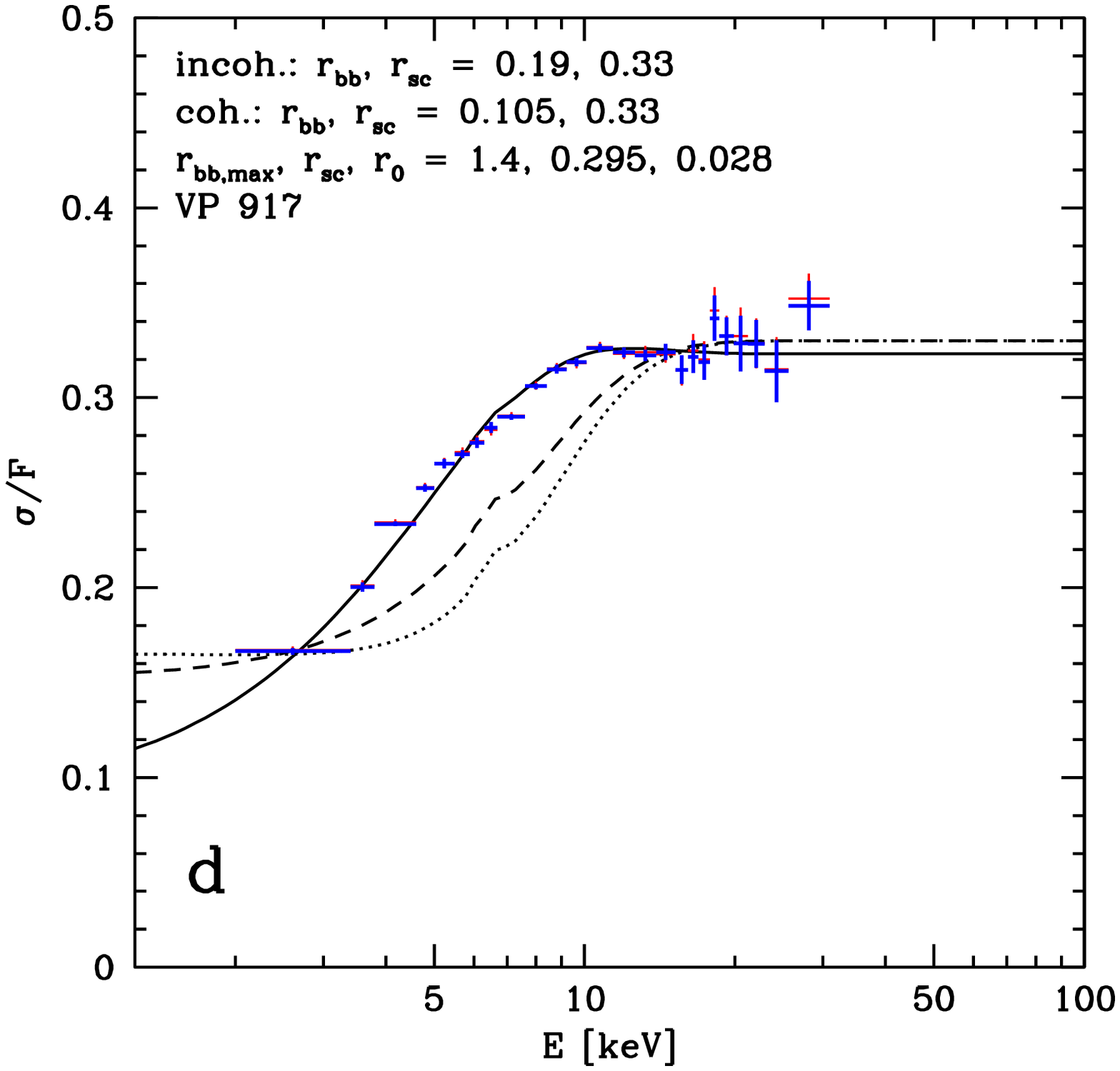,width=7.6cm}}
\caption{The fractional rms variability as a function of energy for the $\chi$-state data. The heavy (blue) and light (red) crosses show the rms results obtained by the direct method and by integrating the PDS, respectively, for the binning of 1/128 s. The dotted and dashed curves show models based on the energy-independent contributions (as fitted by {\sc eqpair}) from unscattered blackbody and scattered photons (including Compton reflection) in the correlated and uncorrelated cases, respectively, and with the parameters given on top of each panel. The solid curves in (a) and in (b), (d)  assume radial variability of the rms in the scattered component and in the blackbody component, respectively, with the parameters given in the 3rd line of each panel. 
\label{rms} }
\end{figure*}

\begin{figure}
\centerline{\psfig{file=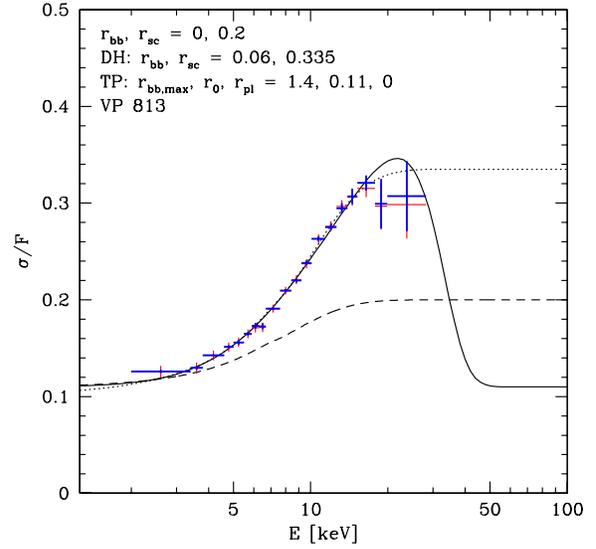,width=7.6cm}}
\caption{The fractional rms variability as a function of energy for the $\omega$ state (VP 813). The data points and the dashed curve have the meaning analogous to that of Fig.\ \ref{rms}. The dotted curve shows a model corresponding to the fit including an additional disc blackbody component (Section \ref{omega}), with the parameters given in the second line. The solid curve shows a model based on the description of the spectrum by a disc blackbody with a local rms varying with radius and a power law, with the parameters shown in the third line above. See Section \ref{frac_m} for details.
\label{813_rms} }
\end{figure}

Figs.\ \ref{rms} and \ref{813_rms} show the results of calculating the rms using the two methods described in Section \ref{data} for the binning of 1/128 s. We see they yield very similar results. The small differences, with the PDS method yielding somewhat lower values of the rms at the lowest energies, are due to the finite range of the frequencies integrated in that method. We note that the shown rms corresponds to most of the total variability power of the source, at least for $f\ga 10^{-3}$ Hz, see Figs.\ \ref{power}, \ref{e_power}, \ref{813_power}. 

We have modelled the rms dependencies by assuming contributions from the three fitted spectral components shown in Fig.\ \ref{chi_models}, i.e., the unscattered blackbody disc emission, Compton scattered component, and Compton reflection. We assign each component a value of the fractional rms, $r_i(E)\equiv \sigma_i(E)/F_i(E)$, which yields the total fractional variability of
\begin{equation}
r_{\rm cor}(E)=\sum_i r_i(E) {F_i(E)\over F(E)},
\label{coh}
\end{equation}
in the case of the variability in the different components being completely correlated (coherent), and
\begin{equation}
r_{\rm unc}(E)^2=\sum_i r_i(E)^2 {F_i(E)^2\over F(E)^2},
\label{incoh}
\end{equation}
in the case of uncorrelated (incoherent) variability. Here $F_{i}(E)$ is the averege flux in the $i$-th spectral component, and $F$ is the total flux. For notational simplicity, we omit hereafter the average sign for $F$, i.e., $F(E)=\langle F(E)\rangle$, $F_i=\langle F_i(E)\rangle$. In general, $r_{\rm cor}\geq r_{\rm unc}$. 

We have tested that allowing the relative variability of the reflection, $r_{\rm refl}$, to be independent of the two other amplitudes, the one for blackbody, $r_{\rm bb}$, and for scattering, $r_{\rm bb}$, only slightly improves the agreement between our models and the data. In fact, we expect the reflection component to respond to the incident flux from the Comptonizing plasma. Thus, we hereafter assume that the two components are fully correlated and $r_{\rm refl}=r_{\rm sc}$.

We then first consider the case of both $r_{\rm bb}$ and $r_{\rm sc}$ independent of energy. We have found this gives a fair description of the rms observed during VP 601 and 720, as shown in Figs.\ \ref{rms}(a), (c). The dashed and dotted curves show the cases of the correlated and uncorrelated variability, respectively. For VP 601, both give similar description of the data. However, those data were obtained in a mode allowing only very broad energy bins below $\sim$15 keV, thus allowing only a rough test of the agreement. On the other hand, the data for VP 720 clearly prefer the correlated variability between the blackbody and scattered photons. (Note, though, that we have not performed here formal $\chi^2$ fitting.) This is also the case for the data sets of VP 619 and 917, where the uncorrelated variability predicts the rms spectrum more below the data than the correlated one. The cause of it is the dip in the rms profile around the intersection of the two components (each dominant at either low or high energies) due to partial cancellation of the two independent variability patterns. The coherence implied by the data (common in accreting black-hole sources, Vaughan \& Nowak 1997) then can be understood by the feedback loop between seed blackbody photons and scattered photons. The Comptonizing plasma upscatters the seed photons, thus reacting to changes in their flux, but also the blackbody photons are partly due to the reprocessing of the emission of the hot plasma directed towards the disc. The presence of reprocessing is implied by the appearance of the strong Compton reflection in the data (see Table 2). 

The model with correlated, energy-independent, variability in the two components gives a marginal description of the VP 619 (though the model curve lies below most of the data points), but it fails for the VP 917, see Figs.\ \ref{rms}(b), (d). The disagreement occurs mostly in the region where the disc blackbody photons dominate the spectrum, see Fig.\ \ref{chi_models}. Still, we find the flattening at the energy at which the unscattered disc blackbody no longer contributes to the spectrum roughly corresponds to the flattening in the observed rms dependence, which reassures us of the correctness of our general spectral decomposition. 

Thus, we have considered the rms of the disc blackbody component being energy-dependent. Physically, the emission in the disc-blackbody spectrum comes from a wide range of radii (Mitsuda et al.\ 1984), and the variability properties are expected to be dependent on the radius as perturbations propagate towards the black hole (Lyubarskii 1997; Kotov et al.\ 2001; \.Zycki 2003). The problem of rms from accretion flows has been studied by Zdziarski (2005, hereafter Z05), who has, in particular, obtained formulae for the rms of the disc blackbody assuming a power law radial dependence of the local rms. We have found that the model for $r_{\rm bb}(E)$ with variability of disc regions with the logarithmic length of $\Delta\! \ln R=0.1$ each varying independently of each other, the value of $kT_{\rm bb}$ as in Table 2, the disc dissipation profile of Shakura \& Sunyaev (1973) assuming the disc extending to the minimum stable orbit, and the radial dependence of the local variability index following the dissipation rate with the power law index of $\beta=4$ [equations (11), (17) and (25) in Z05] gives good description to our data. (We stress again that we have not performed here a formal $\chi^2$ fitting, and certainly there exist other models that can provide similarly good, or better, description.) The models, including the scattering component with the rms independent of energy correlated with the disc blackbody component, are shown by the solid curves in Fig.\ \ref{rms}(b), (d). We have also found an additional energy-independent rms component is required, which can be either correlated or uncorrelated with respect to the blackbody and scattering components. The cases shown in Fig.\ \ref{rms}(b), (d) assume the $E$-independent rms component to be correlated with the other ones.

We then see that the data for the VP 601 show the rms in the scattered component, $\ga$15 keV, to continue to increase. This may be due to a radial dependence for that component, similarly as for the blackbody component. We again used a model of Z05, assuming the local emission hardens from the photon index of 3.5 at 100 times the inner disc radius to 2.5 at the inner radius, the e-folding energy of 300 keV, the fully correlated variability [equations (20--21) and (6) in Z05], and the dissipation profile as above except for $\beta=2$. The resulting model is shown by the solid curve in Fig.\ \ref{rms}. Again, this is just one of many possible models, and the limited rms data at high energies prevent any better constraints. 

The rms data for the VP 813 ($\omega$ state) are shown in Fig.\ \ref{813_rms}. We show here the rms spectrum corresponding to the entire data set, including the dips. However, the dips contribute relatively little to the total rms, and the rms for the high-flux state only looks almost the same. This is also related to the fact that the variability is the strongest in the high energy tail (Fig.\ \ref{813_rms}) in spite of the spectra of the high and low-flux states (Fig.\ \ref{813_eeuf}) differing mostly at low energies. However, the transitions between those flux states occur only relatively infrequently, with most of the variability occuring within the high-flux state.

In our main {\sc eqpair} model either with or without reflection (Table 2), the flux in unscattered blackbody photons is comparable to the flux in scattered photons already at low energies, see Fig.\ \ref{813_model}(a). Thus, assigning the scattered component even a very low energy-independent value already yields a substantial rms at low energies. Consequently, the upper limit on $r_{\rm sc}$ is given by the rms at $\sim$3 keV, and we require $r_{\rm bb}=0$. Such a model is shown by the dashed curve in Fig.\ \ref{813_rms}, and we see it fails completely at $\ga$5 keV. Allowing the unscattered blackbody component alone to have a radial dependence, similarly to the cases of VP 619 and 917, does not improve the fit substantially. We also have considered the case in which the nonthermal electrons have a different variability amplitude than the thermal ones. However, this still does not improve the fit.

On the other hand, we find that our alternative model with the additional disc blackbody component (DH), Fig.\ \ref{813_model}(b) in Section \ref{omega}, does provide a very good description of the rms profile. The dotted curve in Fig.\ \ref{813_rms} shows a very good fit obtained by assuming the variability of the additional disc blackbody and of the hybrid plasma being fully correlated, with the rms of the latter $\gg$ that of the former. 

Since the hybrid plasma in this model covers an inner part of the disc, this model implies the local rms increasing towards the center, with the above two-component model being probably an approximation to a continuous rms radial dependence. In order to test this possibility, we first use the values of $F_i(E)/F(E)$ from the decomposition of the spectrum into thermally-Comptonized disc blackbody  and a power law (model TP in Section \ref{omega}). We then assume the former varies as a disc blackbody with a power-law radial rms dependence (Z05). This is justified by the blackbody fit to the $\la$20 keV spectrum (which yields the maximum temperature of $kT_{\rm bb}=2.3$ keV) approximating it to within 10 per cent. We use the specific rms model of Fig.\ 9 in Z05, which uses the same assumptions (except for $kT_{\rm bb}$) as the models above for VP 619 and 917. The power law tail is assumed to vary independently of the main component. The increasing contribution of the power law at $\ga$20 keV results then in a break in the rms dependence. This model yields a very good description of the observed rms profile, as shown by the solid curve in Fig.\ \ref{813_rms}. We have found, however, that the data do not constrain the rms of the power law component, $r_{\rm pl}$, except for an upper limit of $\la 0.3$. Thus, as a minimum assumption, we set $r_{\rm pl}=0$. To account for the flattening of the rms spectrum at low energies, we also need an energy-independent overall variability, assumed here to be uncorrelated with respect to the blackbody-like component. 

\subsection{Frequency dependence of the rms and energy spectra of the variable components}
\label{qpo}

In Section \ref{frac_m}, we have used the rms integrated over the most (or entire) available frequency ranges, not taking into account the possible dependencies on a chosen frequency range. As seen in Figs.\ \ref{e_power} and \ref{813_power}, the corresponding energy dependencies of the shape of the power spectra are relatively small for the $\chi$ state, and somewhat stronger for the $\omega$ state. Detailed studies of those dependencies are beyond the scope of this paper. However, we do consider here the rms dependence for QPOs (present in the $\chi$ state).

\begin{figure}
\centerline{\psfig{file=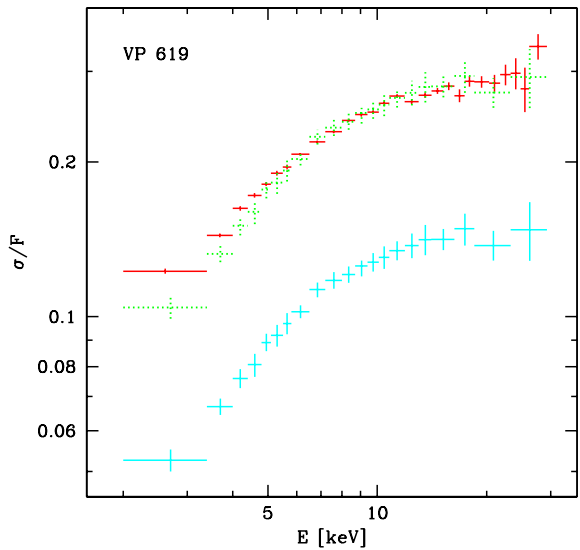,width=7.6cm}}
\caption{The fractional rms variability as a function of energy for both the integrated PDS (the upper solid crosses) and the $\sim$4 Hz QPO (the lower crosses) of the VP 619. The dotted symbols show the rms of the QPO rescaled by a factor of 1.98.
\label{619_qpo} }
\end{figure}

We first considered the VP-619 PCA data. We have fitted its energy-dependent power spectra by sums of Lorentzians (e.g., Nowak 2000; Belloni, Psaltis \& van der Klis 2002), and calculated the QPO power from the amplitude of its Lorentzian. Fig.\ \ref{619_qpo} shows the resulting QPO rms dependence together with the rms integrated over the 1/512--64 Hz range (as in Fig.\ \ref{rms}b). We see the shape of both dependencies are virtually identical at $E\ga 5$ keV, while the QPO rms decreases faster with the decreasing energy than the total rms at $E\la 5$ keV. In the framework of our two-component, blackbody and Comptonization, model, this can be understood as the QPO variability following mostly that of the Comptonizing corona, and much less that of the blackbody.

A general feature of most of the rms dependencies shown in Figs.\ \ref{rms}--\ref{813_rms} is a flattening at $E\ga 15$ keV. This means that the spectrum of the variable component becomes identical to that of the time-average one. On the other hand, the variability is reduced at lower energies due to the presence of a less varying component, presumably the disc blackbody.

\begin{figure}
\centerline{\psfig{file=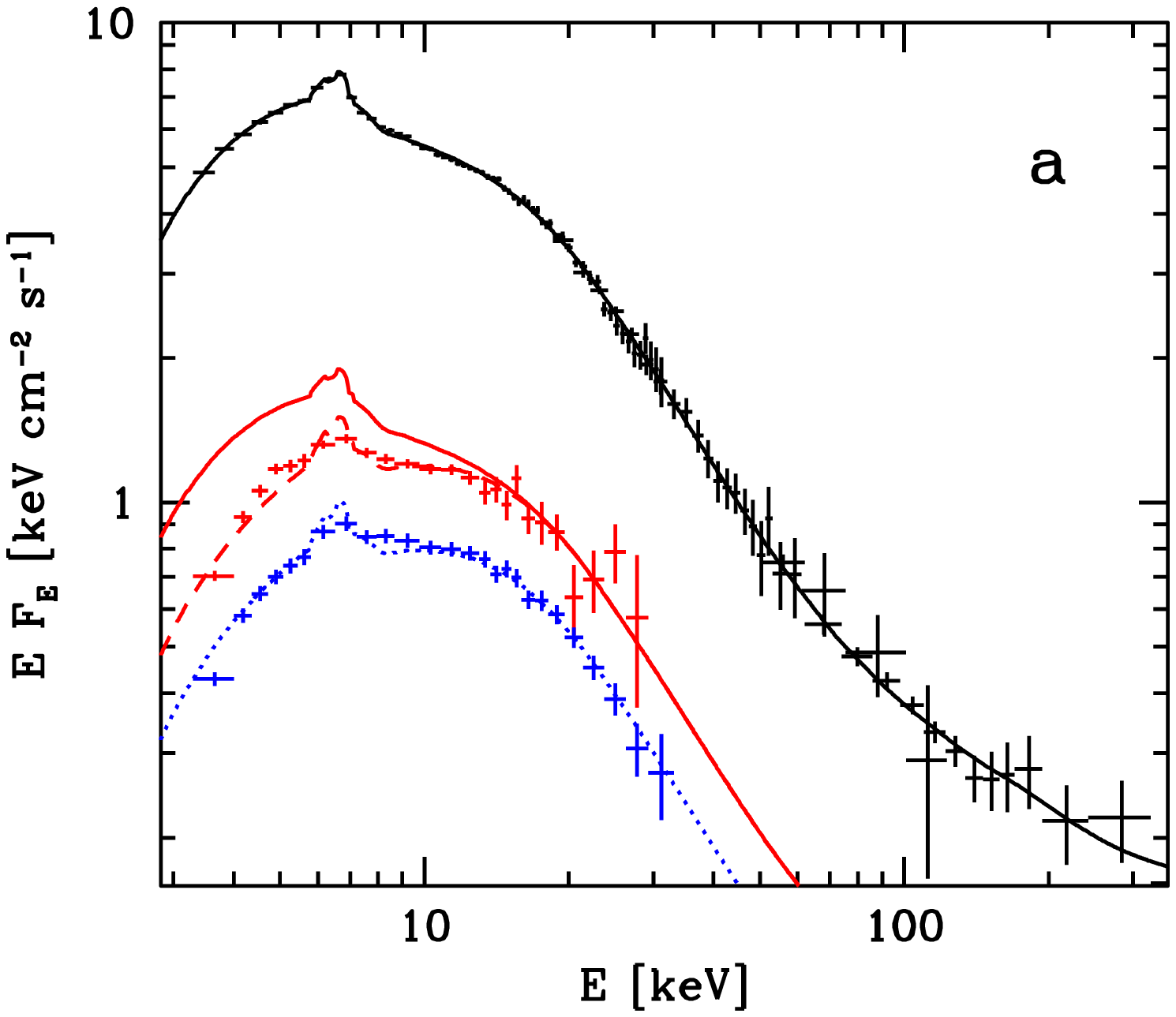,width=7.8cm}}
\centerline{\psfig{file=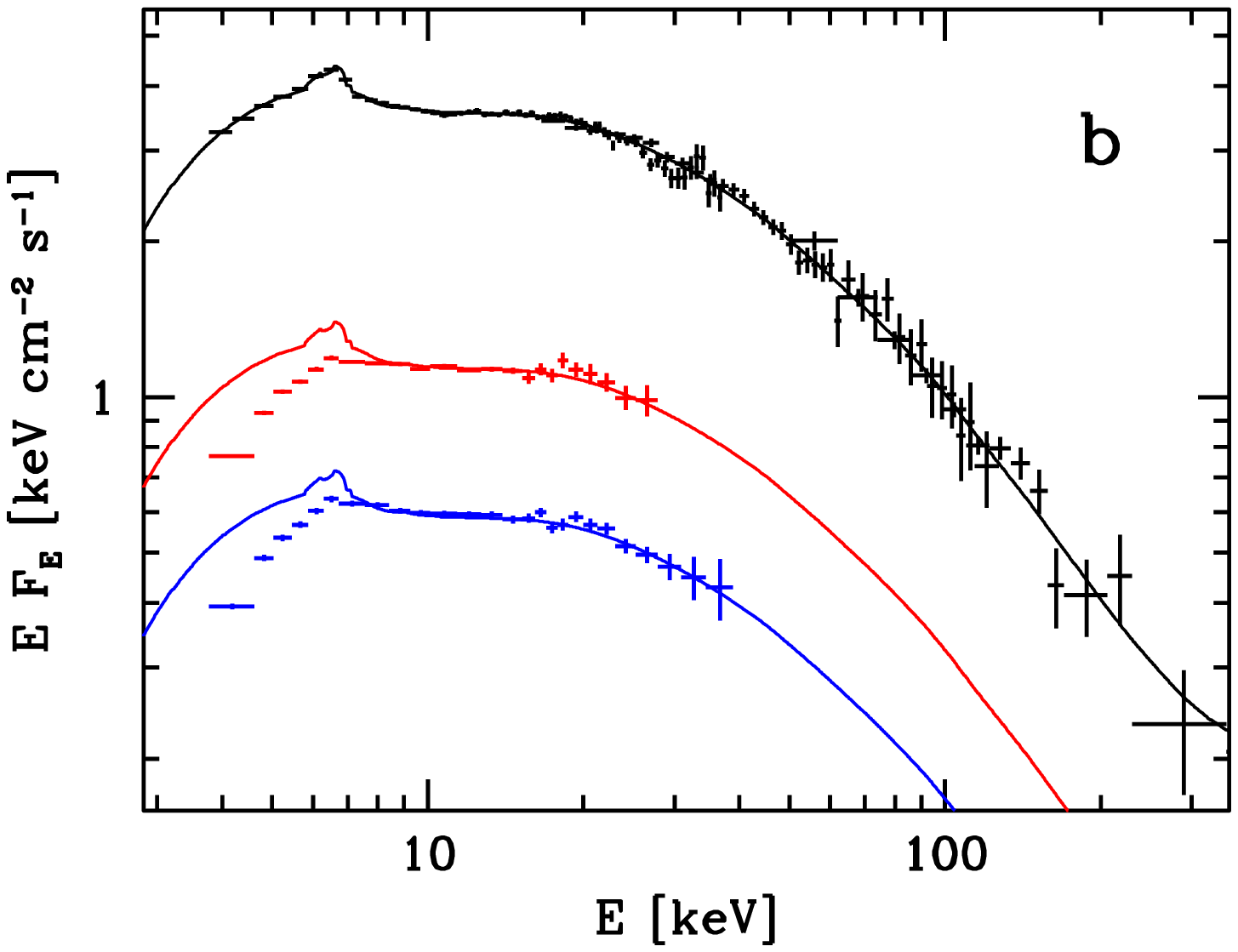,width=7.8cm}}
\caption{The average, total variable, and QPO (from top to bottom) energy spectra for the (a) VP 720 and (b) VP 917. The solid curves give the best-fit models to the average spectra as well as those rescaled to the high-energy parts of the variable spectra. The dashed and dotted curve in (a) give models with a reduced unscattered blackbody component rescaled to the level of the variable and QPO spectra, respectively, see Section \ref{qpo}. The shown spectra have been obtained by fitting the {\sc eqpair} model.
\label{var_spec} }
\end{figure}

This can be also seen from another representation of the energy dependent variability, namely the energy spectrum of a variable component,
\begin{equation}
F_{\rm var}(E)=\sigma(E)= r(E) F(E),
\label{fvar}
\end{equation}
used, e.g., by M\'endez et al.\ (1997) and Gilfanov, Revnivtsev \& Molkov (2003). Fig.\ \ref{var_spec}(a) shows the variable spectra for the VP 720 corresponding to integration of the PDS in the ranges of 1/512--64 Hz, i.e., corresponding to most of the measured variability, and the 1.4--1.9 Hz, containing the QPO. We see both variable spectra have relatively similar shapes, consistent with the result for the VP 619. They are also very similar to the shape of the average spectrum at $E\ga 15$ keV, but fall below that at lower energies. This is again consistent with the picture in which the variable component, including the QPO, arises predominantly in the Comptonizing corona, whereas the blackbody emission is much less variable. The normalization of the 1/512--64 Hz spectrum at $E\ga 10$ keV, as shown by the lower solid curve, is 0.24 with respect to the average spectrum, i.e. equal to that used in the corresponding Fig.\ \ref{rms}(c) for the rms of the Compton component. 

The dashed curve corresponds to the coherent superposition of the unscattered blackbody and Compton components with the relative normalizations as in Fig.\ \ref{rms}(c), i.e., 0.05 and 0.24, respectively. We see that it approximates well the variable spectrum, consistent with the good fit of the rms model in Fig.\ \ref{rms}(c). We also see that the QPO spectrum is very similar to that of the total variable component. It is well approximated by the dotted curve, which has the same shape as th dashed one, but is rescaled by a factor 2/3. Note that the exact shape of the deconvolved spectra in Fig.\ \ref{var_spec} is model-dependent to certain degree.

Fig.\ \ref{var_spec}(b) shows the analogous results for the VP 917. In this case, our fit to the data (Section \ref{spectra}) does not allow a simple two-component modelling of the rms dependence (Fig.\ \ref{rms}d), which may be due to the radial dependence of the rms of the blackbody (Section \ref{frac_m}). Thus, we show here only the total model to the average spectrum renormalized to the levels of the variable components. We still see the same effects as before, i.e., the high-energy, now $E\ga 8$ keV, parts of the variable spectra have the same shape as the average spectrum, and the QPO spectrum is approximately the same as the total variable one. 

We note that recently Rodriguez et al.\ (2004) have arrived at a different conclusion. Namely, they claimed evidence for the presence of an additional component with weak variability at {\it high\/} energies. They based that conclusion on (i) an apparent decrease (fitted by an e-folded power law) of the QPO rms at $E\ga 15$ keV in a PCA observation of \source\ (obs.\ ID P80127-01-03-00) in the $\chi$ state and (ii) the corresponding PCA/HEXTE spectrum brighter at high energies than their fitted absorbed e-folded power law model (fig.\ 3 in Rodriguez et al.\ 2004). 

Given the potential importance of such a result for understanding of the underlying physics of \source, we have repeated their rms analysis for the same observation. With the current PCA calibration, we have found only a flattening of the QPO rms($E$) at high energies, very similar to our results for the VP 619 and 917 above, with no evidence of any decline at highest energies. Also, that dependence is much better fitted by a broken power law (with the best-fit index $\simeq 0$ above the break) than by an e-folded power law (which improvement has the chance probability of 0.009 from the F-test). We also note that the rms decline at high energies was seen by Rodriguez et al.\ (2004, see their fig.\ 4) only in one out of 8 observations. In the remaining 7 observations, a flattening at $\ga 15$ keV was observed, fully consistent with our results above. Furthermore, the e-folded power law used by them to fit the energy spectrum is clearly not a physical model.  That spectrum is very similar to our $\chi$-state spectra (Fig.\ \ref{chi_eeuf}), which are very well fitted by hybrid Comptonization, with no need for any additional component at high energies.

Rodriguez et al.\ (2004) have also suggested that the presumed additional high-energy component originated from the nonthermal synchrotron emission of a jet. We note here that even if there were evidence for such a component, this interpretation has severe problems. Namely, the high-energy part of the spectrum shown in fig.\ 3 of Rodriguez et al.\ (2004) is approximately of a power-law shape with $\Gamma\simeq 3$. Such a steep component when extrapolated to low energies would contain enormous luminosity (as pointed out by Zdziarski et al.\ 2003), as well it would greatly exceed the observed range of the IR flux, see, e.g., fig.\ 6 in Ueda et al.\ (2002). E.g., an extrapolation to $\sim 10^{15}$ Hz, where the nonthermal electrons would no longer cool efficiently (e.g., Heinz 2004), with $\Gamma=3$ and at lower energies with $\Gamma=2.5$ (i.e., with the canonical $\Delta\Gamma$ due to radiative cooling) down to the turnover frequency at $\sim 10^{13}$ Hz yields the luminosity in that component as high as $\sim 10^{42}$ erg s$^{-1}$. A high-energy cutoff in a much harder nonthermal synchrotron spectrum fine-tuned to start at tens of keV would alleviate that problem. However, the OSSE data show no evidence for such a cutoff in the $\chi$ state up to at least $\sim$600 keV (Paper I) or more (Ueda et al.\ 2002). On the other hand, a nonthermal synchrotron cooling break in hard X-ray is highly unlikely (e.g., Heinz 2004). Summarizing, we can rule out the presence of a nonthermal synchrotron component at hard X-rays in the $\chi$ state based on the spectral data alone. 

\subsection{Alternative rms models}
\label{alternative}

An rms increasing with energy can be due to a pivoting process, in which the average spectrum is modulated proportionally to a varying power law fixed at the pivot energy, $E_{\rm p}$, for $E>E_{\rm p}$ (Zdziarski et al.\ 2003, eq.\ [A6]). With $E_{\rm p}\approx 2$ keV, this model with an addition of an $E$-independent variability can roughly (but significantly worse than the best models in Section \ref{frac_m}) describe the increase of the rms at $E\la 10$ keV, but not the subsequent flattening. Also, it predicts the rms increasing with decreasing energy below $E_{\rm p}$. Pivoting occurs, e.g., when a thermally-Comptonizing plasma with an approximately constant power is irradiated by a variable flux of seed soft photons, which is characteristic to the hard state (Zdziarski et al.\ 2002). The pivot energy is then in the middle of the Comptonization spectrum, which is clearly not the case now. Thus, this model appears unlikely for the data presented here.

A pattern of the rms increasing with energy occurs in the soft state of Cyg X-1 on long time scales (Zdziarski et al.\ 2002). Those authors interpreted it as variability of the power supplied to the hybrid plasma at the power in the blackbody seed photons varying much less (Churazov, Gilfanov \& Revnivtsev 2001). This is similar to the models presented in Section \ref{frac_m}. More sophisticated models of this type could take into accout the changes of the spectral shape with the change of the plasma power. This could be modelled in the {\sc eqpair} as variability of $\lh$ at the constant $\ls$ (Zdziarski et al.\ 2002). However, such detailed simulations are beyond the scope of this paper and will be presented elsewhere (Gierli\'nski \& Zdziarski, in preparation). 

Yet another possibility is time evolution of flares. \.Zycki (2004, fig.\ 4), basing on the magnetic flare scenario of Poutanen \& Fabian (1999), has found the rms increasing from 0.12 at 1 keV to 0.21 at 30 keV. This increase is too shallow to account for most of our rms spectra,  but these numbers depend on the model parameters used. On the other hand, it is unclear how this model can explain the flattening of the rms at high energies seen in our data. 

\section{Conclusions}

We confirm the result of Paper I of a very good description of broad-band X-ray/soft \g-ray spectra of \source\ by hybrid Comptonization of disc blackbody emission, by using now all available \xte/OSSE data in the $\chi$ and $\omega$ variability states. In the case of the $\omega$ state, we find that model applies to the spectra of both the high-flux steady emission and the dips. Analyzing the energy as well as power spectra, we identify the $\chi$ state with the intermediate/very high state of black-hole binaries. In the $\omega$ state, we see strong and variable thermal Comptonization of the disc blackbody by a cool, Thomson-thick plasma, followed by a relatively weak high-energy tail.  

Our major new results concern the rms energy dependencies. We find the rms spectra can be well fitted by a coherent superposition of rms contributions from the fitted spectral components, namely the less-variable disc blackbody and more variable Comptonization. Most of our data show a flattening of the rms at high energies, consistent with the domination of the Comptonization spectral component with the rms independent of energy. This is also the case for the rms energy dependence for the QPOs present in the $\chi$ state, which have energy spectra similar to those of the total variable components. On the other hand, some of our data require the rms of the disc blackbody emission to increase with energy, possibly due to the perturbation amplitude in the accretion flow increasing with the decreasing radius. 

\section*{ACKNOWLEDGMENTS}

We thank Juri Poutanen for his valuable comments and suggestions, Piotr \.Zycki for discussion on his theoretical calculations of the rms, and Tomaso Belloni, Marat Gilfanov and Jerome Rodriguez for valuable discussions. This research has been supported by KBN grants 1P03D01827, 1P03D01727, PBZ-KBN-054/P03/2001 and 4T12E04727, and the exchange program between the Polish Academy of Sciences and the Indian National Science Academy.

\label{lastpage}
\end{document}